\documentclass[twocolumn,amsfonts,showpacs,superscriptaddress,nofootinbib]{revtex4-1}
\usepackage{hyperref}
\usepackage[dvipdfmx]{graphicx}
\usepackage{amssymb,amsmath,amsthm,booktabs,mathtools}
\usepackage{bbm}
\usepackage{bm}% bold math
\usepackage{color}

\usepackage{tikz}
\usepackage{pgfplots}
\usepackage{enumitem}

\newcommand{\bra}[1]{{\langle #1 \vert}}

\newcommand{\ket}[1]{{\vert #1 \rangle}}

\newcommand{\x}{{\rm x}}

\def\one{\mathbbm{1}}

\def\be{\begin{equation}}
\def\ee{\end{equation}}

\def\ket#1{|#1\rangle}
\def\bra#1{\langle#1|}

\def\tit#1{}

\usepackage{color}

\usepackage{amsmath}	% required for `\align' (yatex added)
\begin{document}

\title{Operator Entanglement in Interacting Integrable Quantum Systems: \\ the Case of the Rule 54 Chain}

\author{V. Alba}
\affiliation
{Institute  for  Theoretical  Physics, Universiteit van Amsterdam,
Science Park 904, Postbus 94485, 1098 XH Amsterdam,  The  Netherlands}
\author{J. Dubail}
\affiliation
{Laboratoire de Physique et Chimie Th\'eoriques, CNRS, UMR 7019, Universit\'e de Lorraine, 54506 Vandoeuvre-les-Nancy, France
}
\author{M. Medenjak}
\affiliation
{Institut de Physique Th\'eorique Philippe Meyer, \'Ecole Normale Sup\'erieure, \\ PSL University, Sorbonne Universit\'es, CNRS, 75005 Paris, France}

\begin{abstract}
	In a many-body quantum system, local operators in Heisenberg picture $O(t) = e^{i H t} O e^{-i H t}$ spread as time increases. Recent studies have attempted to find features of that spreading which could distinguish between chaotic and integrable dynamics. The operator entanglement --- the entanglement entropy in operator space --- is a natural candidate to provide such a distinction. Indeed, while it is believed that the operator entanglement grows {\it linearly} with time $t$ in {\it chaotic} systems, we present evidence that it grows only {\it logarithmically} in generic interacting {\it integrable} systems. Although this logarithmic growth has been previously established for non-interacting fermions, there has been no progress on interacting integrable systems to date. In this Letter we provide an analytical upper bound on operator entanglement for all local operators in the ``Rule 54'' qubit chain, a  cellular automaton model introduced in the 1990s~[Bobenko {\it et al.}, CMP {\bf 158}, 127 (1993)], and recently advertised as the simplest representative of interacting integrable systems. 
Physically, the logarithmic bound originates from the fact that the dynamics of the models is mapped onto the one of stable quasiparticles that scatter elastically. The possibility of generalizing this scenario to other interacting integrable systems is briefly discussed.
 \end{abstract}

\maketitle

Understanding the out-of-equilibrium dynamics of isolated quantum many-body systems has been a prominent challenge since the early days of quantum mechanics~\cite{neumann1929beweis}. A key recurring idea 
is that, at long times, local properties are captured by statistical ensembles~\cite{neumann1929beweis,rigol2007relaxation,eisert2015quantum,essler2016quench}, despite the 
global dynamics being unitary. This suggests the possibility of a huge compression of information. In one dimension (1d) it implies that the reduced density matrix of a subsystem 
goes to a steady state well approximated by a Matrix Product Operator (MPO) 
\cite{zwolak2004mixed,verstraete2004matrix,hastings2006solving,prosen2007efficiency,
vznidarivc2008complexity,molnar2015approximating}. This contrasts with the intermediate time behavior, 
where one faces an ``entanglement barrier''~\cite{dubail2017entanglement,alba2018quantum} reminiscent of the generic linear growth of the entanglement entropy of a pure state after a quantum quench~\cite{calabrese2005evolution}.

In the late 2000s, the physical intuition that it could sometimes be more efficient to simulate the dynamics of operators ---e.g. density matrices--- rather than the one of pure states spurred another idea~\cite{prosen2007efficiency,prosen2007operator,pizorn2009operator,hartmann2009density,muth2011dynamical}:
that local observables in Heisenberg picture, $O(t) = e^{i H t} O e^{-i H t}$, could also be approximated that way. In an insightful paper, Prosen and \v{Z}nidari\v{c}~\cite{prosen2007efficiency} observed numerically that there was a crucial distinction to be made between {\it chaotic} \cite{D_Alessio_2016,Borgonovi_2016} and {\it non-interacting} dynamics: the bond dimension necessary for an MPO representation of $O(t)$ was apparently blowing up {\it exponentially} with $t$ in the former case and {\it polynomially} in the latter.

%
%
%%%%%%%%%%%%%%%%%%%%%%%%%%%%%%%%%
\begin{figure}[t]
\includegraphics[width=0.98\linewidth]{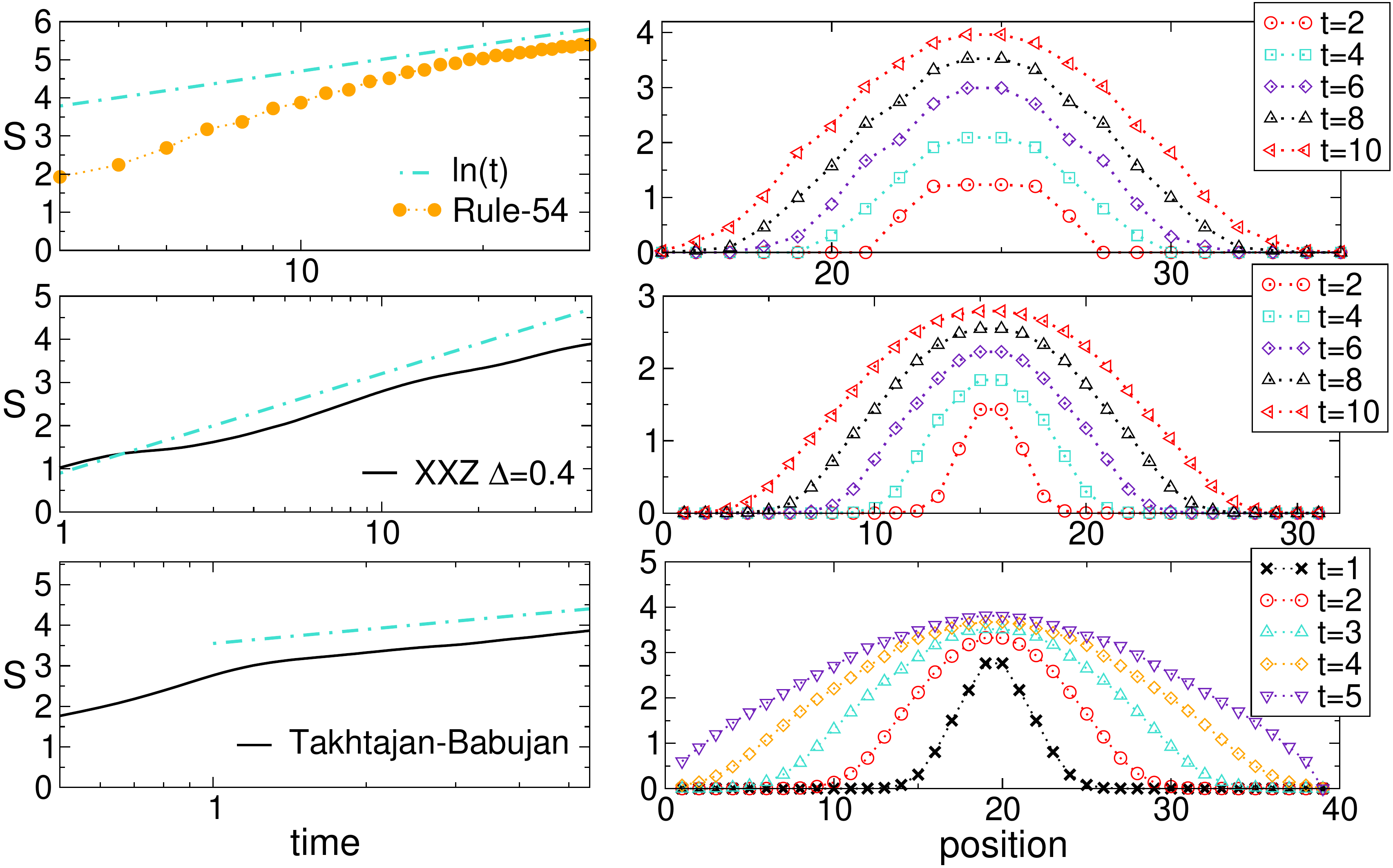} \vspace{-0.2cm}
\caption{Numerical results for the OE $S(O(t))$ for bipartition $A = (-\infty, x]$, $B = (x, \infty)$, in three interacting integrable models. (Left) Growth of $S(O(t))$ at $x=0$: in all three models the growth appears to be logarithmic for local operators: (top) $O = S^+_x$ in the Rule 54 chain, (middle) $O = S^z_x$, $S^+_x$ and  $S_x^zS_{x+1}^z$ in the XXZ chain at $\Delta=0.4$, (bottom) $O= S^z_x$ in the spin-1 Babujan-Takhtajan chain.  
(Right) Profile of the OE for $O=S_x^z$ at different times in the same three models. 
The operator spreading is clearly visible in all cases.}  \vspace{-0.6cm}
	\label{fig1}
\end{figure}
%%%%%%%%%%%%%%%%%%%%%%%%%%%%%%%%%%
%
%

An important figure of merit for the efficiency of this approach is 
the so-called {\it Operator Entanglement} (OE), defined as 
follows. Consider a bipartition of the system $A \cup B$, and the 
Schmidt decomposition of an operator $O$ as $O/\sqrt{\textrm{Tr}(O^\dagger O)}=
\sum_{i} \sqrt{\lambda_i}O_{A,i}\otimes O_{B,i}$, 
where $O_{A,i}$ and $O_{B,i}$ are orthonormal operators, $\textrm{Tr}(O_{A(B),i}^\dagger O_{A(B),j})=\delta_{ij}$, with supports
in $A$ and $B$ respectively, and the Schmidt coefficients $\lambda_i >0$ satisfy the normalization condition $\sum_{i}
\lambda_i = 1$. In complete analogy with state entanglement, one defines the OE as $S(O) \equiv- \sum_i \lambda_i \ln\lambda_i$. 
The OE was first introduced in the context of quantum information~\cite{zanardi2001entanglement} and later connected to MPO-simulability of quantum dynamics~\cite{prosen2007operator,vznidarivc2008complexity,pizorn2009operator,
hartmann2009density,muth2011dynamical,dubail2017entanglement,zhou2017operator}. In the past months, there has been growing interest in the OE, both in condensed matter and in high-energy theory where it connects to quantum chaos, black holes, complexity and models of emergent spacetime~\cite{jonay2018coarse,xu2018accessing,van2018entanglement,pal2018entangling,takayanagi2018holographic,nie2018signature}.

\vspace{0.1cm}\noindent {\bf\em The question.} In this Letter we focus on infinite spin chains with dynamics 
generated by a Hamiltonian $H$ ---or more generally by a unitary evolution 
operator $U$---, and an operator $O$, which has initially  a
finite support located around the origin $x=0$. 
Under time evolution the local operator $O(t) = e^{i H t} O e^{-i H t} $ ---or $U^{-t} O U^t$--- spreads. Like others before us~\cite{prosen2007operator,pizorn2009operator,hartmann2009density,muth2011dynamical,dubail2017entanglement,jonay2018coarse}, we want to understand how $S(O(t))$ grows with time, for the bipartition $A = (-\infty, x]$, $B = (x, \infty)$. In Refs.~\cite{prosen2007operator,pizorn2009operator} it was found numerically that the OE grows at most logarithmically with time in systems with underlying non-interacting fermion dynamics (see Ref.~\cite{dubail2017entanglement} for an analytic derivation), while the behavior of OE in chaotic systems seems to be strikingly different, exhibiting linear growth \cite{jonay2018coarse}. Here, in contrast with these previous works on OE, we focus on the dynamics of interacting integrable systems \cite{Vidmar_2016,Caux_2016,Essler_2016,Ilievski_2016}. The question which motivates us is:

\vspace{0.1cm} {\it Does the growth of $S(O(t))$ distinguish chaotic from interacting integrable dynamics?}

\vspace{0.1cm} We stress that this question is very timely also for a different reason. Operator spreading has been the subject of extremely intense study in chaotic models in the past years, although it is not very clear whether looking simply at the growth of the support of an operator $O(t)$, or equivalently at out-of-time-ordered correlators (OTOC), does reveal any distinctive features of chaos~\cite{khemani2018velocity,gopalakrishnan2018operator,gopalakrishnan2018hydrodynamics} in lattice models with finite-dimensional local Hilbert space like quantum spin chains. For instance, the front of the operator $O(t)$ simply moves ballistically with a diffusive broadening in chaotic~\cite{nahum2018operator,von2018operator,chan2018solution} and integrable~\cite{gopalakrishnan2018operator,gopalakrishnan2018hydrodynamics,PhysRevLett.121.160603} systems alike.
	Therefore it is important to propose new quantities that are truly able to distinguish chaotic from integrable systems. 
	
%The OE of local operators is a very natural candidate for that, provided that the answer to the above question is positive. 

\vspace{0.1cm}\noindent {\bf\em Numerics and general scenario.} An {\it affirmative answer} to the above question is supported by numerical results. In Fig.~\ref{fig1} we display the OE for two well-studied interacting integrable
models (spin-1/2 XXZ and spin-1 Takhtajan-Babujian chains~\cite{takhtajan1982picture,babujian1983exact}). On the accessible time scales, which are relatively short, the results are compatible with $\mathcal{O}(\log t)$ scaling. More importantly, for our purposes, the behavior of the OE appears to be qualitatively the same as the one found in a third interacting integrable
model: the Rule 54 chain (defined below), which is at the center of this Letter. For that particular model, we prove that the OE is (at most) logarithmic for any local operator $O$, thus providing the first indisputable check of the logarithmic growth of OE beyond non-interacting models.

\begin{figure*}
\begin{tikzpicture}
	\draw (-8.7,-1.5) node{\small (a)};
	\draw (-4.6,-1.5) node{\small (b)};
	\draw (-0.1,-1.5) node{\small (c)};
	\draw (2.8,-1.5) node{\small (d)};
	\draw (0,0) node {\includegraphics[width=0.995\textwidth]{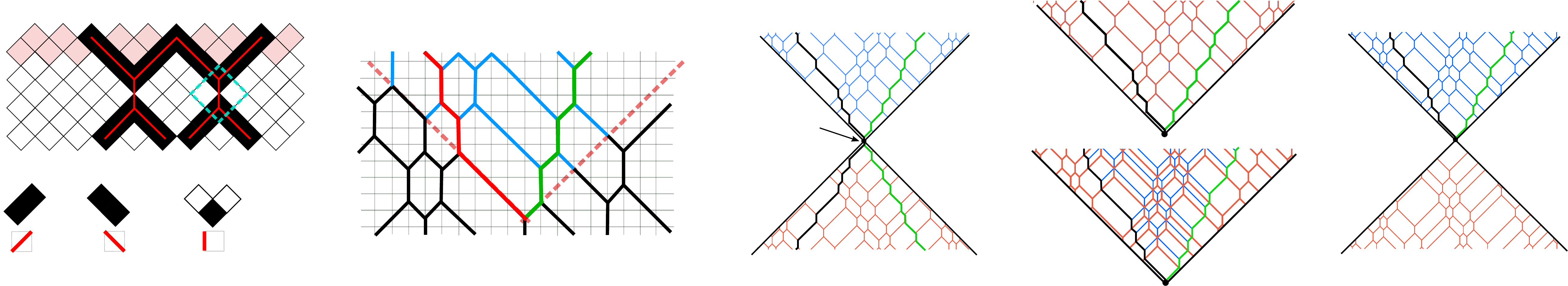}};
	\draw (-7.9,-0.08) node{\tiny -$4$};
	\draw (-7.57,-0.08) node{\tiny -$2$};
	\draw (-7.21,-0.08) node{\tiny $0$};
	\draw (-6.9,-0.08) node{\tiny $2$};
	\draw (-6.59,-0.08) node{\tiny $4$};
	\draw (-6.74,-0.18) node{\tiny $3$};
	\draw (-7.05,-0.18) node{\tiny $1$};
	\draw (-7.4,-0.18) node{\tiny -$1$};
	\draw (-7.73,-0.18) node{\tiny -$3$};
	\draw (-6.11,-0.19) node{\tiny $j$};
	\draw[->] (-5.3,-0.1) -- ++(0,1.4) node[above]{\small $t$};
	\draw (-2.95,-1.2) node{\tiny $0$};
	\draw (-2.76,-1.2) node{\tiny $1$};
	\draw (-2.57,-1.2) node{\tiny $2$};
	\draw (-2.38,-1.2) node{\tiny $3$};
	\draw (-2.19,-1.2) node{\tiny $4$};
	\draw (-3.14,-1.2) node{\tiny -$1$};
	\draw (-3.34,-1.2) node{\tiny -$2$};
	\draw (-3.54,-1.2) node{\tiny -$3$};
	\draw (-0.95,-0.91) node{\tiny $t=0$};
	\draw (-0.95,-0.55) node{\tiny $t=1$};
	\draw (-0.95,-0.18) node{\tiny $t=2$};
	\draw (-0.95,0.19) node{\tiny $t=3$};
	\draw (-0.95,0.57) node{\tiny $t=4$};
	\draw (-0.95,0.95) node{\tiny $t=5$};
	\draw (0.2,0.25) node{\small $O$};	
	\draw[<-] (-2.21,1.1) -- ++(0,0.22) node[above]{\small $x_2$}; 
	\draw[<-] (-4,1.1) -- ++(0,0.22) node[above]{\small $x_1$};
	\draw[very thick,dashed,gray] (6.1,1.7) -- (6.1,0) -- (2.4,0) -- (2.4,-1.7);
	\draw[<-] (0.16,1.35) -- ++(0,0.22) node[above]{\tiny $x_1$};
	\draw[<-] (1.58,1.35) -- ++(0,0.22) node[above]{\tiny $x_2$};
	\draw (0.9,1.3) node[above]{\small $\left| s \right>$};
	\draw (0.9,-1.8) node[above]{\small $\left< s \right|$};
	\draw (4.3,1.6) node[above]{\small $\left| s \right>\left| s \right>$};
	\draw[<-] (6.9,1.35) -- ++(0,0.22) node[above]{\tiny $x_1$};
	\draw[<-] (8.3,1.35) -- ++(0,0.22) node[above]{\tiny $x_2$};
	\draw (7.7,1.3) node[above]{\small $\left| s \right>$};
	\draw (7.7,-1.8) node[above]{\small $\left< s' \right|$};
	\draw (2.9,-1.03) node[above]{\small $\left| s \right>\left| s' \right>$};
	\draw[thick,->,gray] (1.8,0.5) -- ++(1.3,0.3);
	\draw[gray] (2.5,0.6) node[below]{\footnotesize folding};
	\draw[thick,->,gray] (6.8,-0.5) -- ++(-1.2,-0.3);
	\draw[gray] (6.3,-0.1) node[below]{\footnotesize folding};
\end{tikzpicture}
\caption{Operator spreading in the Rule 54 chain. (a) Example of the dynamics generated
by (\ref{eq:U}) acting on the qubit chain, here drawn as a staggered lattice: qubits in the state '1' ('0')
are drawn as black (white) squares. %Shaded boxes denote sites at the same time.
 Red lines superimposed on the black squares show the left and right moving solitons. 
 The dashed box highlights a scattering event, where two solitons get time delayed. The mapping 
 from qubits to solitons is illustrated at the bottom: left and right moving solitons correspond to nearest-neighbor 
 black sites, while a scattering pair corresponds to a single black site surrounded by two white ones. (b) Spacetime
 picture of a typical evolution.
 (c) Spreading of a diagonal operator $O=\left|0 \check{1} 0 \right>\left<0 \check{1} 0 \right|$: the forward and backward lightcones are present. After folding one is back to the situation (c). The solitonic algorithm is able to tell whether a soliton configuration $\left| s\right> \left< s\right|$ contributes or not to $O(t)$. (d) 
 Spreading of off-diagonal operators $O=\left|0 \check{1} 0 \right>\left<0 \check{0} 0 \right|$. Folding the forward and backward lightcones, one sees that the configurations $\left| s\right>, \left| s'\right>$ coincide for $x<x_1$ or $x>x_2$, while $\left| s'\right>$ it can be obtained from $\left| s\right>$ inside the interval $(x_1,x_2)$ simply by applying a time shift of one unit time.
}
\label{fig2}
\end{figure*}
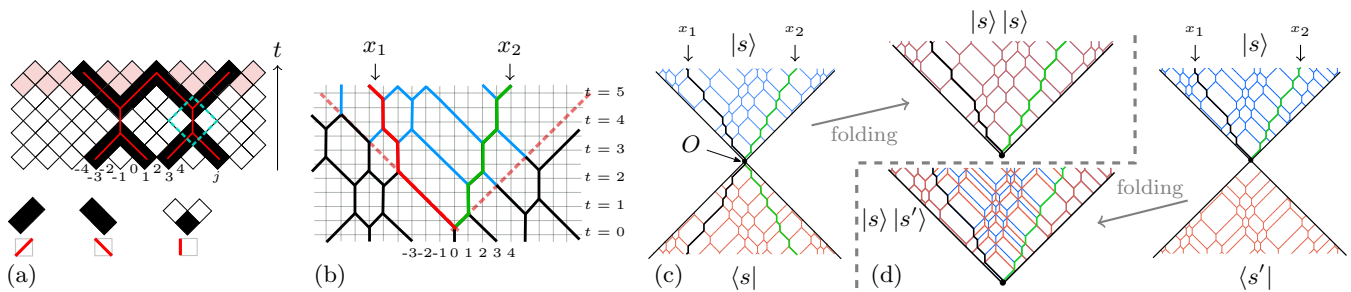

Interestingly, the physical ingredient that 
	underlies our result is the presence of infinite-lifetime excitations 
	(solitons) that undergo two-body elastic scattering during  the 
	evolution of the operator $O(t)$ (see 
	Fig.~\ref{fig2}). In contrast, in a chaotic system the operator $O$ 
	will generate excitations that will propagate, eventually decay 
	and then create more excitations. This causes  
	any memory of the initial infinite temperature state (the identity) 
	to be lost in an expanding region around $x=0$. Our findings 
	in the Rule 54 chain suggest a totally different scenario in the 
	integrable case. There, $O$ generates only a few {\it stable} 
	excitations that propagate ballistically through the 
	system. They still affect the initial infinite temperature state 
	in an expanding region around $x=0$, 
	but in a much less dramatic way. The stable excitations emitted 
	by $O$ simply shift 
	the positions of the other ones as they 
	scatter with them (Fig.~\ref{fig2}). 
	Then the full dynamics of $O(t)$ is accurately 
	reconstructed from the knowledge of the number of those 
	scatterings.

\vspace{0.1cm}
\noindent {\bf\em The Rule 54 chain.} We focus on the Rule 54 qubit chain~\cite{bobenko1993two}, a model studied recently in Refs.~\cite{prosen2016integrability,prosen2017exact,gopalakrishnan2018facilitated,gopalakrishnan2018operator,klobas2018time,gopalakrishnan2018hydrodynamics,buvca2019exact} ---it has also been named ``Toffoli-gate model''~\cite{gopalakrishnan2018facilitated} or ``Floquet-Fredrickson-Andersen model''~ \cite{gopalakrishnan2018operator,gopalakrishnan2018hydrodynamics} in relation with other recent work~\cite{rowlands2018noisy}---.
It has been establishing itself as the simplest model exhibiting generic physical properties of interacting integrable systems \cite{bobenko1993two,gopalakrishnan2018operator,klobas2018time,gopalakrishnan2018hydrodynamics}. These range from the coexistence of ballistic and diffusive transport \cite{klobas2018time} to the generic behavior of the OTOC front \cite{gopalakrishnan2018hydrodynamics}, which are absent in non-interacting systems \cite{Spohn_2018,Khemani_2018}. The key microscopic feature which distinguishes interacting from non-interacting integrable dynamics is the {\it time delay} associated with scattering events between pairs of stable excitations. While excitations are either delayed or hastened as they scatter in interacting systems, this is not the case in free models where particles remain unaffected. The Rule 54 chain is a genuine interacting model because it has a non-zero time delay (Fig.~\ref{fig2}). Its dynamics has all the salient features of soliton gases or ``flea gas'' models~\cite{boldrighini1983one,doyon2018soliton,bulchandani2018bethe} which correctly reproduce the large-distance and long-time behavior of out-of-equilibrium integrable systems~\cite{doyon2018geometric,cao2018incomplete,gopalakrishnan2018operator,gopalakrishnan2018hydrodynamics,caux2017hydrodynamics}.

% {\color{red} These results imply that there is a significant difference between the behavior of free and interacting integrable systems, which we observe also in the behavior of OE. Indeed, interacting and non-interacting integrable models seem to differ by a prefactor in-front of the logarithm. XXX Jerome: maybe it's dangerous to start talking about the prefactor, since it is an important quantity yet we do not have a good understanding of it. So I would propose to remove these two sentences XXX}

%Indeed, the dynamics of the model is the one of a very simple classical gas of solitons, and this allows for an analytical treatment which goes well beyond anything that was known previously for other interacting systems \cite{bobenko1993two,klobas2018time}.

The Hilbert space of the model $\mathcal{H} \equiv (\mathbb{C}^2)^\mathbb{Z}$ corresponds to an infinite chain of qubits, with a dynamics generated locally by a unitary gate $U_{x}$ acting on sites $x-1$, $x$ and $x+1$ ($x \in \mathbb{Z}$) as
\begin{eqnarray}
		\label{eq:gate}
\nonumber		U_{x} &=& \left| 1 0 1 \right> \left< 1 1 1 \right|  +    \left| 1 0 0 \right> \left< 1 10 \right|  + \left| 1 1 1 \right> \left< 101 \right|   \\
\nonumber					&& + \left| 1 1 0 \right> \left< 1 0 0 \right|  +    \left| 0 0 1 \right> \left< 0 1 1 \right|  + \left| 0 1 0 \right> \left< 0 1 0 \right|  \\
					&& +  \left| 0 1 1 \right> \left< 0 0 1 \right|  +    \left| 0 0 0 \right> \left< 0 0 0 \right|   .
\end{eqnarray}
The gate updates the central qubit $x$, depending on the state of the two adjacent ones.  The name ``Rule 54'', introduced by Wolfram in the context of cellular automata~\cite{wolfram1983statistical}, stems from the binary encoding `$00110110$' of the number 54, which corresponds to the outcoming state of the central qubit in each of the eight terms in Eq.~\eqref{eq:gate}. Time evolution is generated by
	\begin{equation}
		\label{eq:U}
		U \equiv  ( \prod_{x \, {\rm even}} U_{x} ) \times (  \prod_{x \, {\rm odd}} U_{x} ).
	\end{equation}
The dynamics defined by (\ref{eq:U}) sustains left- and right-moving 
solitons with constant velocity, which get time-delayed by a single unit of time when they scatter (Fig.~\ref{fig2}). We find it convenient to introduce an operator $M: (\mathbb{C}^2)^{\mathbb{Z}} \rightarrow (\mathbb{C}^2)^{\mathbb{Z}\cup (\mathbb{Z}+\frac{1}{2})}$ that transforms qubit configurations into {\it soliton} ones. The latter live on the lattice $\mathbb{Z} \cup (\mathbb{Z} + \frac{1}{2})$ comprising integer and half-integer sites. $M$ is defined by the two following rules. The half-integer site $x+\frac{1}{2}$ is occupied by a soliton iff both qubits $x$ and $x+1$ are in state `$1$'. The integer site $x$ is occupied by a pair of scattering solitons iff spins $x-1$, $x$, $x+1$ are in the configuration `$010$'.

We stress that, even though the evolution generated by $U$ simply maps one computational basis state to another, to study operator spreading one has to expand the initial operator in the computational basis, which results in a non-trivial (quantum) superposition~\cite{gopalakrishnan2018operator,gopalakrishnan2018hydrodynamics}.

\vspace{0.1cm}
\noindent {\bf\em Upper bound on OE.} Before we delve deeper into the Rule 54 chain, let us stress the following simple fact about OE: if the operator $O$ can be decomposed in the form $O=\sum_{i \in I} \tilde{O}_{A,i} 
\otimes \tilde{O}_{B,i}$, then its OE is automatically bounded,
\begin{equation}
\label{ub}
S(O)\leq \log |I|,
\end{equation}
where $ |I| $ is the number of terms in the sum.

The bound follows from the definition of OE (see above). Indeed, when the terms in the sum are orthonormal, 
$\textrm{Tr}(\tilde{O}_{A(B),i}^\dagger \tilde{O}_{A(B),j})=\delta_{ij}$, it is clear that $S(O) = \log |I|$. Instead, if this is not the case, one can always decompose the operator $O$ with respect to the two orthonormal sets $\{ O'_{A,i} \}_{i \in I}$ and $
\{ O'_{B,j} \}_{j \in I}$ obtained from a Schmidt orthogonalization of  $\{ \tilde{O}_{A,i} \}_{i \in I}$ and $\{ \tilde{O}_{B,i} \}_{i \in I}$, in the form $O/\sqrt{{\rm Tr}(O^\dagger O)}=\sum_{i,j \in I} \Lambda_{ij} O'_{A,i}
\otimes O'_{B,j}$. Making a singular value decomposition of the matrix $ [\Lambda]_{i,j}=\Lambda_{ij} $, $\Lambda = U^\dagger \cdot {\rm diag}(\sqrt{\lambda_1}, \sqrt{\lambda_2}, \dots ) \cdot V$, one obtains the Schmidt decomposition $O/\sqrt{{\rm Tr}( O^\dagger O)} = \sum_{i \in I} \sqrt{\lambda_i} O_{A,i} \otimes O_{B,i}$, with orthonormal sets $O_{A,i} = \sum_j U_{ij} O'_{Aj}$, $O_{B,i} = \sum_j V_{ij} O'_{Bj}$. This yields the OE $S(O) = - \sum_{i\in I}  \lambda_i \log \lambda_i \leq \log |I|$, with the bound saturated only if all the $\lambda_i$'s are identical.

\vspace{0.1cm}
\noindent {\bf\em The solitonic algorithm.} \; The upper bound for the rule 54 spin chain is rooted in the existence of an algorithm which decides whether or not a given pair of solitons at time $ t $ emerged from the origin (adapted from \cite{klobas2018time}). [The reader is invited to practice the algorithm with the example in Fig. \ref{fig2}(b) with $t=5$, $x_1= -\frac{11}{2}$, $x_2=\frac{7}{2}$.]

Consider a configuration with a left mover at $x_1$ (either a single soliton or a scattering pair) and a right mover at $x_2$. We want to know if they both came from $x=0$ at $t=0$. The algorithm uses two counters $j_l$, $j_r$, initialized as $j_l= -2t + \frac{1}{2}$ , $j_r = x_2+ ((-x_2- \frac{1}{2} ) \; {\rm mod}\; 2)$. It reads the configuration site by site, from right to left, starting at site $ x_2-\frac{1}{2} $. If a site is unoccupied, the counters remain unchanged; if a site is occupied by a left(right)-mover, their values change as $j_r \rightarrow j_r +2$ ($j_l \rightarrow j_l + 2$). A scattering pair counts for both a left and a right mover, so both counters must be updated. The algorithm stops when it arrives at site $ x_1 $. At this point the value of the two counters is checked: the pair at $x_1$, $x_2$ came from the origin iff $j_l = x_1 - ((x_1 - \frac{1}{2}) \; {\rm mod}\; 2)$ and $j_r = 2 t-\frac{1}{2}$.%We call this the halting condition.

The crucial point is that, since both counters remain in the interval $[-2t,2t]$, the set of internal states $ \ket{j_l,j_r} $ explored by the algorithm is a subset of $I \equiv [-2t,2t]^2$, which is of size $| I | =  \mathcal{O}(t^2) $.
%\begin{enumerate}
%	\item independent of the soliton configuration
%	\item of size $| I | =  \mathcal{O}(t^2) $.
%\end{enumerate}

 \vspace{0.1cm} 
%A similar bound holds for operators initially supported on $l>1$ sites because they are products of one-site ones. 
%The result for one-site operators implies that for operators 
%with $l>1$ the number of terms is at most $\mathcal{O}(t^{2 l})$. 

 \vspace{0.1cm}
\noindent {\bf\em The operator decomposition.}\, For simplicity, we study the OE in the soliton basis.
The scaling of the OE with time is the same as in the qubit basis, because the linear map $M$ between the qubit and soliton basis is local. By ``local'' we mean that there exists a decomposition $ M=\sum_{i,j=1}^\chi M_{A,i}\otimes M_{B,j} $ where $ \chi $ is finite and time independent \cite{SM_M}.  %for any bipartition, with $ A $ and $ B $ connected.
This implies $M^\dagger \left(\sum_{i \in I} \tilde{O}_{A,i} \otimes \tilde{O}_{B,i}\right) M  =\sum_{ i\in I} \sum_{j,k = 1,\dots, \chi} M_{A,j}^\dagger\tilde{O}_{A,i} M_{A,k} \otimes M_{B,j}^\dagger \tilde{O}_{B,i} M_{B,k}$, so a decomposition with $|I|$ terms in the soliton basis implies a similar decomposition with $\chi^2\times  |I| $ terms in the qubit basis. Since $\chi$ is constant,the scaling with $t$ is unchanged because it only depends on how fast the set $I$ grows.
%\begin{eqnarray}
%& &M^\dagger \left(\sum_{i \in I} \tilde{O}_{A,i} \otimes \tilde{O}_{B,i}\right) M \\
%&& =\sum_{\underset{ j,k= 1,\dots, \chi}{i\in I}} M_{A,j}^\dagger\tilde{O}_{A,i} M_{A,k} \otimes M_{B,j}^\dagger \tilde{O}_{B,i} M_{B,k}. \nonumber
%\end{eqnarray}

We consider two sets of operators separately. The first one corresponding to the diagonal operators and the second one to the non-diagonal operators.

\vspace{0.1cm} 
\noindent{\em Diagonal operators.} It is sufficient to provide a bound on the OE for the projector $\ket{1} \bra{1}$ (in the computational basis), since the other projector $\ket{0} \bra{0}$ is obtained by subtracting it from the identity.

The way the operator $\ket{1}\bra{1}$ acts on solitons is most easily seen by expanding the identities on the two neighboring sites, namely (a 'check' designates the qubit at site $j=0$)
\begin{eqnarray}
	\label{eq:diag4}
	\left| 1  \right> \left< 1\right|  & = & \left| 0\check{1}0  \right> \left< 0\check{1}0 \right| +   \left| 0\check{1}1  \right> \left< 0\check{1}1 \right| \qquad \\
\nonumber		&& +   \left| 1\check{1}0  \right> \left< 1\check{1}0 \right| +   \left| 1\check{1}1  \right> \left< 1\check{1}1 \right|.
\end{eqnarray}
The term $M \left| 0\check{1}0  \right> \left< 0\check{1}0 \right| M^\dagger$ is the 
projector on configurations with a pair of scattering solitons emerging from 
the origin at $t=0$. The second (third) term projects on 
configurations with a single left (right) mover at the origin, and the 
fourth term is a projector on configurations with left and right 
movers emitted simultaneously from the origin. For simplicity we 
discuss only the time evolution of the first term $D(t) \equiv M U^{-t} 
\left| 0\check{1}0  \right> \left< 0\check{1}0 \right| U^t M^\dagger$. The other 
three terms can be treated in a similar way~\cite{SMdiagonal}.

We focus on an entanglement cut in the middle of the chain ---i.e. a bipartition $A = (-\infty,0], B=(0,+\infty)$--- where $S(O(t))$ is maximal (Fig.~\ref{fig1}). For a fixed time $t >0$, we define a set of projectors $\mathcal{A}^{x}_{j_l, j_r}$ acting on soliton configurations $\left| s \right>$ as follows. For $x_2 > 0$, $\one_A \otimes\mathcal{A}^{x_2}_{j_l, j_r}  \left| s \right> = \left|  s\right>$ if the site $x_2$ is occupied by a right mover and if the solitonic algorithm initiated at $x_2$ is in the internal state $\left| j_l, j_r \right>$ when it arrives at the origin ---we stress that we now stop the algorithm when it arrives at the origin, not when it arrives at $j_l$---, and $\one_A\otimes \mathcal{A}^{x_2}_{j_l, j_r}  \left| s \right> = 0$ otherwise. This means that $\mathcal{A}^{x_2}_{j_l,j_r}$ identifies 
the soliton configurations in $A$ that are compatible with the dynamics of the model and 
with the condition that the right mover emitted from the center is at $x_2$ at time $t$. For $x_1<0$, $\mathcal{A}^{x_1}_{j_l, j_r}$ is defined in the same way, but the site $x_1$ has to be occupied by a left mover.

Clearly, since at $t=0$ there is only one right mover emitted from 
$x=0$, for a given $\left| s\right>$ and given 
counters $j_l, j_r$, there can be no more than one value of $x_2>0$ 
for which $\one_A \otimes\mathcal{A}^{x_2}_{j_l, j_r} \left| s \right> = \left| s\right>$ (similarly, no more than one value of $x_1 <0$ for which $\mathcal{A}^{x_1}_{j_l, j_r} \otimes \one_B \left| s \right> = \left| s\right>$). Since the solitonic algorithm detects all configurations at time $t$ which had a pair of solitons at the origin at time $0$, the time evolution of $ D(t) $ can be decomposed as
\begin{equation}
	\label{eq:proj0}
	D(t)= \sum_{(j_l, j_r) \in I} (\sum_{x_1\leq 0} \mathcal{A}_{j_l,j_r}^{x_1} )\otimes( \sum_{x_2>0} \mathcal{A}_{j_l,j_r}^{ x_2} ).
\end{equation}
The decomposition $(\ref{eq:proj0})$ has no more than $|I| = \mathcal{O}(t^2)$ terms in the sum. Taking into account the inequality (\ref{ub}), we see that the OE grows at-most logarithmically. 

Alternatively, the decomposition of the diagonal terms can be obtained from the results of Ref.~\cite{klobas2018time}, which focused on classical dynamics in the Rule 54 chain. This is a consequence of the dynamics generated by (\ref{eq:gate}) preserving the diagonal structure of the operators: for diagonal operators, the mapping $ \ket{s}\bra{s}\to \ket{s} $ actually reduces quantum dynamics to classical dynamics.

%\comJ{[An alternative proof consists in constructing explicitly an MPO representation for $D(t)$ along the lines of Ref.~\cite{klobas2018time}, see the SM~\cite{SMdiagonal}.]}

	\vspace{0.1cm}
\noindent {\em Non-diagonal operators.}\quad It is sufficient to consider the non-diagonal operator $\left| 1 \right> \left< 0 \right|$. In order to express it in terms of solitons, it should be expanded as follows,
\begin{eqnarray}
	\label{eq:nondiag}
\nonumber && \left| 1 \right> \left< 0 \right| \, = \, \left| 0\check{1}0 \right>\left< 0\check{0}0 \right|  \, + \,  \left| 0\check{1}10 \right>\left< 0\check{0}10 \right| 
   \, + \,  \left| 0\check{1}11 \right>\left< 0\check{0}11 \right| \\ 
\nonumber	  && +  \left| 01\check{1}0 \right>\left< 01\check{0}0 \right| \, + \,   \left| 11\check{1}0 \right>\left< 11\check{0}0 \right|  \, + \, \left| 01\check{1}10 \right>\left< 01\check{0}10 \right| +  \\
 && \left| 01\check{1}11 \right>\left< 01\check{0}11 \right|   +  \left| 11\check{1}10 \right>\left< 11\check{0}10 \right| +  \left| 11\check{1}11 \right>\left< 11\check{0}11 \right|  \hspace{-0.08cm}.
\end{eqnarray}
The first term takes a configuration with no soliton at the origin, and creates a pair of solitons; the second term takes a scattering pair on site $x=1$ and replaces it by a single left-moving soliton at $x = \frac{1}{2}$, etc. A detailed study of all nine terms ---which can all be treated in a similar way--- is given in the Supplemental Material~\cite{SMnondiagonal}. Here, for simplicity, we focus only on the first one, $N(t) \equiv M U^{-t}\left| 0\check{1}0 \right>\left< 0\check{0}0 \right| U^t M^\dagger$.
%Once again, due to the deterministic nature of the evolution, all of the terms $\left| s \right> \left< s' \right|$ contributing to $N(t)$ have equal amplitude $ 1 $. 

%Contrary to the diagonal case, the soliton configurations $\ket{s}$ and $\bra{s'}$ can now be different. A typical configuration contributing to $N(t)$, with a pair of solitons created at the origin, is shown in Fig.~\ref{fig2}(d). The key observation is that $\ket{s}$ and $\bra{s'}$ are closely related. Outside the region $[x_1,x_2]$ enclosed by the two solitons coming from the origin, $\ket{s}$ and $\bra{s'}$ are identical, while inside that region, the configuration $\bra{s'}$ (in red in Fig.~\ref{fig2}(d)) is {\it time-shifted} by one time unit compared to $\ket{s}$ (in blue in Fig.~\ref{fig2}(d)).

The main observation (Fig.~\ref{fig2}(d)) is that $N(t)$ is a sum of soliton configurations of the form $\sum_s \ket{s}\bra{s'}$. The sum runs over configurations with a pair of solitons coming from the center, and the  configuration $ \bra{s'} $  is obtained from $ \bra{s} $ by a set of local linear maps $ W_{x_1,x_2} $.
$ W_{x_1,x_2} $ maps $ \bra{s} $ onto $ \bra{s'}$  by erasing the two solitons emerging from the center, and by undoing the effects they had on the remaining solitons (Fig.~\ref{fig2}(d)). To elaborate, let $x_1$ and $x_2$ be the positions of those two solitons at time $t$. Then the linear operator $W_{x_1,x_2}$  annihilates the solitons at $x_1$, $x_2$,  it applies two layers of unitary gates $U_x$ inside the interval $[x_1,x_2]$, and it acts as the identity outside. Because these are local operations, $W_{x_1,x_2}$ itself possesses a decomposition of the form $W_{x_1,x_2} = \sum_{a=1}^{\chi_W} W_{x_1,a}\otimes W_{x_2,a}$, with $\chi_W$ finite, and with operators $ W_{x_1,a} $ and $W_{x_2,a}$ acting on $ A $ and $ B $ respectively. 
%[In other words, $W_{x_1,x_2}$ is an exact MPO with finite bond dimension $\chi_W$, see the SM for details~\cite{SMnondiagonal}.] 
The time evolution of the first non-diagonal term $ N(t) $ can thus be decomposed as
	\begin{equation}
		N(t) =   \sum_{\underset{a = 1,\dots, \chi_W}{(j_l, j_r) \in I} } ( \sum_{x_1 \leq 0} \mathcal{A}_{j_l,j_r}^{x_1}    W_{x_1,a}^\dagger )\otimes ( \sum_{x_2>0}  \mathcal{A}_{j_l,j_r}^{ x_2} W_{x_2,a}^\dagger )  .
	\end{equation}
That decomposition again involves $\chi_W \times |I | = \mathcal{O}(t^2)$ terms, proving the at-most logarithmic growth also for the first non-diagonal term in~(\ref{eq:nondiag}). Similar decompositions can be found for the other eight terms~\cite{SMnondiagonal}.

Importantly, at most logarithmic growth of OE for operators acting nontrivially on a single site, implies logarithmic growth of OE for all local operators.

\vspace{0.1cm}
\noindent {\bf\em Discussion and Conclusion.}\; We have shown that the OE of local operators in Heisenberg picture grows at most logarithmically in the Rule 54 chain. We stress that the two basic ingredients leading to that conclusion are
\begin{enumerate}[label=(\alph*)]
	\item the existence of a quasi-local mapping $M$, which transforms the evolution operator $U$ of the interacting integrable model into the one of a soliton gas $M U M^\dagger$
	\item within that soliton gas, the existence of a solitonic algorithm which can efficiently decide, for any configuration $s$ at time $t$, whether a given soliton at position $x_1$ ---or, as above, two solitons at $x_1$ and $x_2$--- came from the origin at time $t=0$.
\end{enumerate}
	It is tempting to generalize this scenario to other interacting integrable models, in order to get a general theoretical explanation for the logarithmic growth of OE (Fig.~\ref{fig1}). Several recent works point to the validity of (a) for more general interacting integrable models~\cite{doyon2018geometric,doyon2018soliton,cao2018incomplete,bulchandani2018bethe,gopalakrishnan2018operator,gopalakrishnan2018hydrodynamics,caux2017hydrodynamics}. Making that claim more quantitative, and trying to construct such a quasi-local mapping $M$ for, say, the Lieb-Liniger model or the XXZ chain, is a challenging open problem. It seems natural to expect that a mapping 
$M$ exists at least in an approximate sense ---this is in fact underlying the entanglement dynamics after global quenches in integrable 
systems~\cite{AlCal,AlCal-1}---. Then, given a certain soliton gas, for instance the one constructed in Ref.~\cite{doyon2018soliton}, finding an algorithm (b) seems to be a well posed problem; this is an exciting direction for future work.

\vspace{0.1cm}
\begin{acknowledgments}
We thank B. Bertini, K. Klobas, M. Kormos, A. De Luca, A. Nahum, L. Piroli, T. Prosen, and J.-M. St\'ephan for useful discussions and comments on the manuscript. We are grateful to the organizers of the conference ``Non-equilibrium behaviour of isolated classical and quantum systems'' in SISSA, where this work was initiated, for providing a very stimulating environment. Part of this work was supported by the CNRS ``D\'efi Infiniti'' (JD).
\end{acknowledgments}

\bibliography{OE54}

\newpage
%\appendix

	\pagebreak

	\begin{widetext}
		
		\begin{center}
			\textbf{\large Supplementary material:\\ Operator Entanglement in Interacting Integrable Quantum Systems:\\ the Case of the Rule 54 Chain}
		\end{center}

		\section{Operator decomposition and Matrix product operators (MPO)}
		
		The central property in deriving the bound on entanglement is the existence of decomposition
		\begin{equation}
		\label{decomp}
		O=\sum_{i \in I} \tilde{O}_{A,i} 
		\otimes \tilde{O}_{B,i}.
		\end{equation}
		for certain set of operators, where either  $ |I|\propto t^2 $, or $ |I|\propto \mathcal{O}(1) $. In this section we will make a connection between the internal dimension of the MPO and the operator decomposition. 
		
		The tensor associated with the site $ j $ can be identified with a diagrammatic representation
		
				\begin{equation*}
		\begin{tikzpicture}
		\begin{scope}
		\draw (-0.5,0) node[left]{\footnotesize $a'$} -- (0.5,0) node[right]{\footnotesize $a$};
		\draw (0,-0.5) node[below]{\footnotesize $\bra{\sigma}$}-- (0,0.5) node[above]{\footnotesize $\ket{\sigma'}$};
		\filldraw[red] (-0.3,-0.3) rectangle (0.3,0.3);
		\draw (0,0) node{$j$};
		\draw (-3.5,0) node{$\mathcal{A}[j]^{\sigma'_j,a'}_{\sigma_j,a} =$};
		\end{scope}
		\end{tikzpicture}.
		\end{equation*}
		This tensor corresponds to the operator $ \ket{\sigma'}\bra{\sigma} $ on site $ j $, with components $ a' $ and $ a $ on the auxiliary space. The operator acting on the full chain can then be composed by contracting the auxiliary state operators associated with different sites
		\begin{equation*}
		\begin{tikzpicture}
		\draw (0,2) node{$O=\sum_{\sigma_{\{j\}},\sigma_{\{j\}}'}\dots \mathcal{A}[-2]_{\sigma_{-2}}^{\sigma_{-2}'} \mathcal{A}[-\frac{3}{2}]_{\sigma_{-\frac{3}{2}}}^{\sigma_{-\frac{3}{2}}'}  \mathcal{A}[-1]_{\sigma_{-1}}^{\sigma_{-1}'} \mathcal{A}[-\frac{1}{2}]_{\sigma_{-\frac{1}{2}}}^{\sigma_{-\frac{1}{2}}'} \mathcal{A}[0]_{\sigma_0}^{\sigma_0'}  \mathcal{A}[\frac{1}{2}]_{\sigma_{\frac{1}{2}}}^{\sigma_{\frac{1}{2}}'} \mathcal{A}[1]_{\sigma_1}^{\sigma_1'} \mathcal{A}[\frac{3}{2}]_{\sigma_{\frac{3}{2}}}^{\sigma_{\frac{3}{2}}'}  \mathcal{A}[2]_{\sigma_{2}}^{\sigma_{2}'}   \dots $};
		\draw (-6,0) node{$=\sum_{\sigma_{\{j\}},\sigma_{\{j\}}'}$};
		\draw[dashed] (-5,0) -- (-4.3,0);
		\draw[dashed] (5,0) -- (4.3,0);
		\begin{scope}
		\draw (-4.5,0) -- (4.5,0);
		\foreach \x in{-2,-1,...,2} { \draw (2*\x,-0.5) -- (2*\x,0.5) -- (2*\x,-0.5) node[below]{\footnotesize $\bra{\sigma_{\x}}$}-- (2*\x,0.5) node[above]{\footnotesize $\ket{\sigma'_{\x}}$};}
		\foreach \x in{-3,-1,...,3} { \draw (\x,-0.5) -- (\x,0.5) -- (\x,-0.5) node[below]{\footnotesize $\bra{\sigma_{\frac{\x}{2}}}$}-- (\x,0.5) node[above]{\footnotesize $\ket{\sigma'_{\frac{\x}{2}}}$};}
		\foreach \x in {-4,-3,...,4} {
		\filldraw[red] (-0.3+\x,-0.3) rectangle (0.3+\x,0.3) ;
		}
		\foreach \x in{-2,-1,...,2} \draw (2*\x,0) node{$\x$};
		\foreach \x in{-3,-1,...,3} \draw (\x,0) node{$\frac{\x}{2}$};
		\end{scope}
		\end{tikzpicture},
		\end{equation*}
		where the connected legs correspond to the contraction. A decomposition of the operator can now be simply obtained by contracting MPO in the region $ A=(-\infty,0] $ and in the region $ B=(0,\infty)$, and associating the value of the index $ a $ corresponding to the intersection, with an operator $ \bar{O}_{A,a} $
		\begin{equation*}
		\bar{O}_{A,a}=\sum_{\sigma_{\{j\}},\sigma_{\{j\}}'}\dots \mathcal{A}[-2]_{\sigma_{-2}}^{\sigma_{-2}'} \mathcal{A}[-\frac{3}{2}]_{\sigma_{-\frac{3}{2}}}^{\sigma_{-\frac{3}{2}}'}  \mathcal{A}[-1]_{\sigma_{-1}}^{\sigma_{-1}'} \mathcal{A}[-\frac{1}{2}]_{\sigma_{-\frac{1}{2}}}^{\sigma_{-\frac{1}{2}}'} \mathcal{A}[0]_{\sigma_0,a}^{\sigma_0'}.
		\end{equation*}
		Similarly, we can associate the non-contracted index with the operator on the sub-lattice $ B $
		
				\begin{equation*}
		\bar{O}_{B,a}=\sum_{\sigma_{\{j\}},\sigma_{\{j\}}'}  \mathcal{A}[\frac{1}{2}]_{\sigma_{\frac{1}{2}}}^{\sigma_{\frac{1}{2}}',a} \mathcal{A}[1]_{\sigma_1}^{\sigma_1'} \mathcal{A}[\frac{3}{2}]_{\sigma_{\frac{3}{2}}}^{\sigma_{\frac{3}{2}}'}  \mathcal{A}[2]_{\sigma_{2}}^{\sigma_{2}'}   \dots,
		\end{equation*}
		which provides the decomposition \eqref{decomp}. The dimensionality of MPO is therefore directly connected to the number of terms in this decomposition and subsequently to the upper bound on OE. In what follows we provide an explicit MPO representations of the operators presented in the main text. This gives an explicit prescription of how to construct the operators acting on the sublattices introduced in the main text, and proves their existence.
		\section{The mapping $M$ and its formulation as a finite MPO}
		
		The Hilbert space of the qubit chain is $(\mathbb{C}^2)^{\otimes L}$. For convenience, $L$ is assumed to be large but finite. The sites are labeled from $-(L-1)/2$ to $(L-1)/2$, assuming $L$ odd. Like in the main text, we draw the chain as follows (here for $L=11$):
		$$
		\begin{tikzpicture}
		\draw[dashed, thick] (-5,0) -- (-4,1) -- (-3,0) -- (-2,1) -- (-1,0) -- (0,1) -- (1,0) -- (2,1) -- (3,0) -- (4,1) -- (5,0);
		\filldraw (-5,0) circle (0.8mm) node[below]{$-5$};
		\filldraw (-4,1) circle (0.8mm) node[above]{$-4$};
		\filldraw (-3,0) circle (0.8mm) node[below]{$-3$};
		\filldraw (-2,1) circle (0.8mm) node[above]{$-2$};
		\filldraw (-1,0) circle (0.8mm) node[below]{$-1$};
		\filldraw (0,1) circle (0.8mm) node[above]{$0$};
		\filldraw (1,0) circle (0.8mm) node[below]{$1$};
		\filldraw (2,1) circle (0.8mm) node[above]{$2$};
		\filldraw (3,0) circle (0.8mm) node[below]{$3$};
		\filldraw (4,1) circle (0.8mm) node[above]{$4$};
		\filldraw (5,0) circle (0.8mm) node[below]{$5$};
		\end{tikzpicture}
		$$
		We work with the following boundary conditions: we assume that there are `ghost qubits' at sites $-(L+1)/2$ and $(L+1)/2$ which are both in the state `0' and are never updated.
		
		The solitons live on the integer and half-integer sites between $-(L-1)/2$ and $(L-1)/2$:
		\begin{itemize}
			\item left movers live on half-integer sites $j = 2p + \frac{1}{2}$ ($p \in \mathbb{Z}$)
			\item right movers live on half-integer sites $j = 2p - \frac{1}{2}$ ($p \in \mathbb{Z}$)
			\item pairs of scattering solitons live on integer sites $j  \in \mathbb{Z}$.
		\end{itemize}
		Thus, on each integer or half-integer site, we have a local Hilbert space spanned by two states $\left| True \right>$, $\left| False \right>$ (or $\left| T\right>$, $\left| F\right>$) that indicate whether or not the site is occupied. The operator $M$ that maps qubits to solitons (see the main text) can be written as an MPO with bond dimension $4$. The non-zero components of the tensors that enter the MPO all have equal weight $1$. They are drawn as follows:
		$$
		\begin{tikzpicture}[scale=0.4]
		\draw (-5,0) node{$j\in \mathbb{Z}$:};
		\draw (-5,-5) node{$j\in \mathbb{Z}+\frac{1}{2}$:};
		\begin{scope}
		\draw (-1,0) node[left]{\footnotesize 0} -- (1,0) node[right]{\footnotesize 0};
		\draw (0,-1) node[below]{\footnotesize $\left| 0 \right>$} -- (0,1) node[above]{\footnotesize $\left| F \right>$};
		\filldraw[green] (-0.5,-0.5) rectangle (0.5,0.5);
		\draw (0,0) node {$j$};
		\end{scope}
		\begin{scope}[xshift=5cm]
		\draw (-1,0) node[left]{\footnotesize 0} -- (1,0) node[right]{\footnotesize 1};
		\draw (0,-1) node[below]{\footnotesize $\left| 0 \right>$} -- (0,1) node[above]{\footnotesize $\left| F \right>$};
		\filldraw[green] (-0.5,-0.5) rectangle (0.5,0.5);
		\draw (0,0) node {$j$};
		\end{scope}
		\begin{scope}[xshift=10cm]
		\draw (-1,0) node[left]{\footnotesize 1} -- (1,0) node[right]{\footnotesize 2};
		\draw (0,-1) node[below]{\footnotesize $\left| 1 \right>$} -- (0,1) node[above]{\footnotesize $\left| T \right>$};
		\filldraw[green] (-0.5,-0.5) rectangle (0.5,0.5);
		\draw (0,0) node {$j$};
		\end{scope}
		\begin{scope}[xshift=15cm]
		\draw (-1,0) node[left]{\footnotesize 3} -- (1,0) node[right]{\footnotesize 2};
		\draw (0,-1) node[below]{\footnotesize $\left| 1 \right>$} -- (0,1) node[above]{\footnotesize $\left| F \right>$};
		\filldraw[green] (-0.5,-0.5) rectangle (0.5,0.5);
		\draw (0,0) node {$j$};
		\end{scope}
		\begin{scope}[xshift=20cm]
		\draw (-1,0) node[left]{\footnotesize 2} -- (1,0) node[right]{\footnotesize 0};
		\draw (0,-1) node[below]{\footnotesize $\left| 0 \right>$} -- (0,1) node[above]{\footnotesize $\left| F \right>$};
		\filldraw[green] (-0.5,-0.5) rectangle (0.5,0.5);
		\draw (0,0) node {$j$};
		\end{scope}
		\begin{scope}[xshift=25cm]
		\draw (-1,0) node[left]{\footnotesize 2} -- (1,0) node[right]{\footnotesize 1};
		\draw (0,-1) node[below]{\footnotesize $\left| 0 \right>$} -- (0,1) node[above]{\footnotesize $\left| F \right>$};
		\filldraw[green] (-0.5,-0.5) rectangle (0.5,0.5);
		\draw (0,0) node {$j$};
		\end{scope}
		\begin{scope}[xshift=30cm]
		\draw (-1,0) node[left]{\footnotesize 1} -- (1,0) node[right]{\footnotesize 3};
		\draw (0,-1) node[below]{\footnotesize $\left| 1 \right>$} -- (0,1) node[above]{\footnotesize $\left| F \right>$};
		\filldraw[green] (-0.5,-0.5) rectangle (0.5,0.5);
		\draw (0,0) node {$j$};
		\end{scope}
		\begin{scope}[xshift=35cm]
		\draw (-1,0) node[left]{\footnotesize 3} -- (1,0) node[right]{\footnotesize 3};
		\draw (0,-1) node[below]{\footnotesize $\left| 1 \right>$} -- (0,1) node[above]{\footnotesize $\left| F \right>$};
		\filldraw[green] (-0.5,-0.5) rectangle (0.5,0.5);
		\draw (0,0) node {$j$};
		\end{scope}
		\begin{scope}[xshift=0cm, yshift=-5cm]
		\draw (-1,0) node[left]{\footnotesize 0} -- (1,0) node[right]{\footnotesize 0};
		\draw (0,0)-- (0,1) node[above]{\footnotesize $\left| F \right>$};
		\filldraw[green] (-0.8,0.5) -- (0.8,0.5) -- (0,-0.8);
		\draw (0,0) node {$j$};
		\end{scope}
		\begin{scope}[xshift=5cm, yshift=-5cm]
		\draw (-1,0) node[left]{\footnotesize 1} -- (1,0) node[right]{\footnotesize 1};
		\draw (0,0)-- (0,1) node[above]{\footnotesize $\left| F \right>$};
		\filldraw[green] (-0.8,0.5) -- (0.8,0.5) -- (0,-0.8);
		\draw (0,0) node {$j$};
		\end{scope}
		\begin{scope}[xshift=10cm, yshift=-5cm]
		\draw (-1,0) node[left]{\footnotesize 2} -- (1,0) node[right]{\footnotesize 2};
		\draw (0,0)-- (0,1) node[above]{\footnotesize $\left| F \right>$};
		\filldraw[green] (-0.8,0.5) -- (0.8,0.5) -- (0,-0.8);
		\draw (0,0) node {$j$};
		\end{scope}
		\begin{scope}[xshift=15cm, yshift=-5cm]
		\draw (-1,0) node[left]{\footnotesize 3} -- (1,0) node[right]{\footnotesize 3};
		\draw (0,0)-- (0,1) node[above]{\footnotesize $\left| T \right>$};
		\filldraw[green] (-0.8,0.5) -- (0.8,0.5) -- (0,-0.8);
		\draw (0,0) node {$j$};
		\end{scope}
		\end{tikzpicture}
		$$
		and they are contracted in the following way to give an operator that acts on the above qubit chain:
		$$
		\begin{tikzpicture}[scale=0.4]
		\draw (-19.5,0) node{$M \,= $};
		\draw (-16,0) node[left]{\footnotesize 0+1} -- (16,0) node[right]{\footnotesize 0+2};
		\foreach \x in {-5,-3,...,5} \draw (3*\x,-4) -- ++(0,5);
		\foreach \x in {-4,-2,...,4} \draw (3*\x,-2) -- ++(0,3);
		\foreach \x in {-5,-4,...,5}
		{
			\filldraw[green] (-0.5+3*\x,-0.5) rectangle (0.5+3*\x,0.5);
			\draw (3*\x,0) node {\footnotesize $\x$};
		}
		\foreach \x in {-9,-7,...,9}
		{
			\draw (1.5*\x,0) -- ++(0,1);
			\filldraw[green] (-0.8+1.5*\x,0.5) -- (0.8+1.5*\x,0.5) -- (1.5*\x,-0.8);
			\draw (1.5*\x,0) node {\footnotesize $\frac{\x}{2}$};
		}
		\end{tikzpicture}
		$$
		Here the indices '0+1' and '0+2' on the left and right stand for the left and right vectors which enter the MPO. It is easy to check that the components of the MPO are constructed in order to implement the two rules given in the main text.

		Notice that $M^\dagger M = 1$ (the identity on the qubit chain), while $M M^\dagger = 1_{{\rm adm.\; conf.}}$ is the orthogonal projector onto the subspace spanned by all admissible soliton configurations.

		\newpage
		
		\section{More details on the solitonic algorithm}
		We present the solitonic algorithm in further details; we describe it in pseudo-code. The soliton configuration is encoded in the form of a boolean array $\tt s[j]$, with label ${\tt j} \in \mathbb{Z} \cup (\mathbb{Z} + \frac{1}{2})$. $\tt s[j] = True$ means that the site ${\tt j}$ is occupied by a soliton or a pair of scattering solitons, $\tt s[j] = False$ means it is empty. Notice that the information about the 'species' ---namely whether the site is occupied by a pair or by a single soliton, and whether the latter is a right mover or a left mover--- is given by the parity of ${\tt j}$:
		\begin{itemize}
			\item left movers live on half-integer sites ${\tt j} = 2p + \frac{1}{2}$ ($p \in \mathbb{Z}$)
			\item right movers live on half-integer sites ${\tt j} = 2p - \frac{1}{2}$ ($p \in \mathbb{Z}$)
			\item pairs of scattering solitons splitting at time $t$ live on integer sites ${\tt j} = 2p$ ($p \in \mathbb{Z}$)
			\item pairs of scattering solitons fusing at time $t$ live on integer sites ${\tt j} = 2p+1$ ($p \in \mathbb{Z}$)
		\end{itemize}
		The algorithm takes a configuration ${\tt s}$, an integer ${\tt t}$, and two integer or half-integer labels ${\tt x_1}$, ${\tt x_2}$ as an input.
		It determines whether ${\tt s}$ is a configuration at time $t={\tt t}$ which had a pair of particles scattering at the origin $j=0$ at time $t=0$, and if the positions of the left- and right-mover in that pair are ${\tt x_1}$ and ${\tt x_2}$ at time $t$. It works as follows:
		\begin{center}
			\fbox{\begin{minipage}{40em}
					{\tt \#initialize the counters jl and jr\\
						if s[x2] and ((2*x2)\%4=3):    \phantom{aaa}   \#there is a right mover at x2 \\
						\phantom{aaaa}			jl $\leftarrow$ -2*t + 0.5  \\
						\phantom{aaaa}			jr $\leftarrow$ x2 \\
						elseif s[x2] and ((2*x2)\%4=0):      \#there is a (splitting) pair at x2 \\
						\phantom{aaaa}			jl $\leftarrow$ -2*t + 0.5  \\
						\phantom{aaaa}			jr $\leftarrow$ x2+1.5 \\
						elseif s[x2] and ((2*x2)\%4=2):      \#there is a (fusing) pair at x2 \\
						\phantom{aaaa}			jl $\leftarrow$ -2*t + 0.5  \\
						\phantom{aaaa}			jr $\leftarrow$ x2+0.5 \\
						else:	 return False		\phantom{aaaaaaaaaaa}		        \#no right moving soliton at x2, stop here\\

						\#read configuration s from x2 to x1\\
						j $\leftarrow$ x2-0.5 \\
						while j>x1: \\
						\phantom{aaaa}			if s[j] and ((2*s2)\%4=1):  \phantom{aaa} \#left mover at j\\
						\phantom{aaaaaaaa}			jr $\leftarrow$ jr+2 \\
						\phantom{aaaa}			elseif s[j] and ((2*s2)\%4=3):   \#right mover at j\\
						\phantom{aaaaaaaa}			jl $\leftarrow$ jl+2 \\
						\phantom{aaaa}			elseif s[j] and ((2*s2)\%2=0):   \#scattering pair at j\\
						\phantom{aaaaaaaa}			jl $\leftarrow$ jl+2 \\
						\phantom{aaaaaaaa}			jr $\leftarrow$ jr+2 \\
						
						\#check counters \\
						if s[x1] and ((2*x1)\%4=1):	\phantom{aaa}  \#there is a left mover at x1 \\
						\phantom{aaaa} if (jr=2*t-0.5) and (jl=x1): return True \\
						elseif s[x1] and ((2*x1)\%4=0):      \#there is a (splitting) pair at x2 \\
						\phantom{aaaa} if (jr=2*t-0.5) and (jl=x1-1.5): return True \\
						elseif s[x1] and ((2*x1)\%4=2):      \#there is a (fusing) pair at x2 \\
						\phantom{aaaa} if (jr=2*t-0.5) and (jl=x1-0.5): return True \\
						
						\#if "True" not  returned yet, then configuration not valid \\
						return False
					}
			\end{minipage}}
		\end{center}
		As explained in the main text, the key point about this algorithm is that {\it the set of internal states $\left| j_l, j_r \right>$ that are explored is of order $O(t^2)$ at most}. This is clear because both $j_l$ and $j_r$ are (half-)integers between $-2t$ and $2t$.
		
		\newpage

		\section{Details on the diagonal case}
		\label{sec:app:diag}
		Here we give all the details about the four cases in Eq. (3) in the main text. We give all the components that enter the construction of the MPO. The components are represented as follows:
		\begin{equation*}
		\begin{tikzpicture}
		\begin{scope}
		\draw (-0.5,0) node[left]{\footnotesize $(a',j_l',j_r')$} -- (0.5,0) node[right]{\footnotesize $(a,j_l,j_r)$};
		\draw (0,-0.5) node[below]{\footnotesize $\sigma$}-- (0,0.5) node[above]{\footnotesize $\sigma'$};
		\filldraw[red] (-0.3,-0.3) rectangle (0.3,0.3);
		\draw (0,0) node{$j$};
		\draw (-3.5,0) node{$\mathcal{A}[j]^{\sigma',(a',j_l',j_r')}_{\sigma,(a,j_l,j_r)} =$};
		\end{scope}
		\end{tikzpicture}
		\end{equation*}
		and they are contracted as
		\begin{equation*}
		\begin{tikzpicture}
		\draw (0,1) node{$\dots \mathcal{A}[-2]_{\sigma_{-2}}^{\sigma_{-2}'} \mathcal{A}[-\frac{3}{2}]_{\sigma_{-\frac{3}{2}}}^{\sigma_{-\frac{3}{2}}'}  \mathcal{A}[-1]_{\sigma_{-1}}^{\sigma_{-1}'} \mathcal{A}[-\frac{1}{2}]_{\sigma_{-\frac{1}{2}}}^{\sigma_{-\frac{1}{2}}'} \mathcal{A}[0]_{\sigma_0}^{\sigma_0'}  \mathcal{A}[\frac{1}{2}]_{\sigma_{\frac{1}{2}}}^{\sigma_{\frac{1}{2}}'} \mathcal{A}[1]_{\sigma_1}^{\sigma_1'} \mathcal{A}[\frac{3}{2}]_{\sigma_{\frac{3}{2}}}^{\sigma_{\frac{3}{2}}'}  \mathcal{A}[2]_{\sigma_{2}}^{\sigma_{2}'}   \dots $};
		\draw (-6,0) node{$=$};
		\draw[dashed] (-5,0) -- (-4.3,0);
		\draw[dashed] (5,0) -- (4.3,0);
		\begin{scope}
		\draw (-4.5,0) -- (4.5,0);
		\foreach \x in {-4,-3,...,4} {
			\draw (\x,-0.5) -- (\x,0.5);
			\filldraw[red] (-0.3+\x,-0.3) rectangle (0.3+\x,0.3);
		}
		\foreach \x in{-2,-1,...,2} \draw (2*\x,0) node{$\x$};
		\foreach \x in{-3,-1,...,3} \draw (\x,0) node{$\frac{\x}{2}$};
		\end{scope}
		\end{tikzpicture}
		\end{equation*}
		to give a linear operator acting on the space of solitons, which sends the boolean configuration $(\dots \sigma_{-2} \sigma_{-\frac{3}{2}}   \sigma_{-1} \sigma_{-\frac{1}{2}}  \sigma_0   \sigma_{\frac{1}{2}} \sigma_1   \sigma_{\frac{3}{2}}   \sigma_{2} \dots )$ to $(\dots \sigma'_{-2} \sigma'_{-\frac{3}{2}}   \sigma'_{-1} \sigma'_{-\frac{1}{2}}  \sigma'_0   \sigma'_{\frac{1}{2}} \sigma'_1   \sigma'_{\frac{3}{2}}   \sigma'_{2} \dots )$.

		\subsection{Components for first term in Eq. (3): $M U^{-t} \left| 0\check{1}0 \right> \left<0 \check{1} 0 \right| U^t M^\dagger$}	
		\label{sec:appdiag1}
		
		We now list all non-zero components. We have
		\begin{eqnarray*}
			&\qquad &	\mathcal{A}[j]_{\sigma, (0,0,0)}^{\sigma, (0,0,0)}  = \mathcal{A}[j]_{\sigma, (2,0,0)}^{\sigma, (2,0,0)} =  1 , \qquad \sigma = True, False .
		\end{eqnarray*}
		%Similarly, for $j < -2t$, they are
		%\begin{eqnarray*}
		%	&\qquad &	\mathcal{A}[j]_{\sigma, (2,0,0)}^{\sigma, (2,0,0)}  =  1 , \qquad \sigma = True, False . \\
		%\end{eqnarray*}
		This ensures that the MPO acts as the identity outside the light-cone. For $-2t \leq j \leq 2t$, the non-zero components are chosen in order to implement the solitonic algorithm. The 'activation index' $a$ goes from $0$ to $1$ when the right mover coming from the origin is met:
		\begin{eqnarray*}
			j = 2p - \frac{1}{2}, \; p \in \mathbb{Z}: &\qquad &	\mathcal{A}[j]^{True, (1,-2t + \frac{1}{2}, j)}_{True, (0,0,0)}  = 1  \\
			j = 2p+1, \; p \in \mathbb{Z}: &\qquad &	\mathcal{A}[j]^{True, (1,-2t + \frac{1}{2}, j+\frac{1}{2})}_{True, (0,0,0)}  = 1  \\
			j = 2p, \; p \in \mathbb{Z}: &\qquad &	\mathcal{A}[j]^{True, (1,-2t + \frac{1}{2}, j+\frac{3}{2})}_{True, (0,0,0)}  = 1  .
		\end{eqnarray*}
		Similarly, it goes from $1$ to $2$ when the left mover coming from the origin is met:
		\begin{eqnarray*}
			j = 2p + \frac{1}{2}, \; p \in \mathbb{Z}: &\qquad &	\mathcal{A}[j]^{True, (2,0,0)}_{True, (1, j,2t - \frac{1}{2})}  = 1  \\
			j = 2p+1, \; p \in \mathbb{Z}: &\qquad &	\mathcal{A}[j]^{True, (2,0,0)}_{True, (1, j-\frac{1}{2},2t - \frac{1}{2})}  = 1  \\
			j = 2p, \; p \in \mathbb{Z}: &\qquad &		\mathcal{A}[j]^{True, (2,0,0)}_{True, (1, j-\frac{3}{2},2t - \frac{1}{2})}  = 1 .
		\end{eqnarray*}
		Inside the region enclosed by the left and right solitons coming from the origin, the activation index is always $1$:
		\begin{eqnarray*}
			j = 2p + \frac{1}{2}, \; p \in \mathbb{Z}: &\qquad &	\mathcal{A}[j]_{True, (1,j_l,j_r)}^{True, (1, j_l, j_r+2)}  = 1  \\
			j = 2p - \frac{1}{2}, \; p \in \mathbb{Z}: &\qquad &	\mathcal{A}[j]_{True, (1,j_l,j_r)}^{True, (1, j_l+2, j_r)}  = 1  \\
			j  \in \mathbb{Z}: &\qquad &	\mathcal{A}[j]_{True, (1,j_l,j_r)}^{True, (1, j_l+2, j_r+2)}  = 1  \\
			\forall \,	j  : &\qquad &	\mathcal{A}[j]_{False, (1,j_l,j_r)}^{False, (1, j_l, j_r)}  = 1  .
		\end{eqnarray*}
		%FInally, at all empty sites inside the region enclosed by the two solitons we have
		%$$
		%\mathcal{A}[j]_{False, (1,j_l,j_r)}^{False, (1, j_l, j_r)}  = 1 .
		%$$

		\subsection{Second term: $M U^{-t} \left| 0\check{1}1 \right> \left<0 \check{1} 1 \right| U^t M^\dagger$}	
		\label{sec:appdiag2}
		
		Now we need to make sure that there is a single left-moving soliton at position $j=\frac{1}{2}$ at time $t=0$, instead of the pair of scattering solitons that we had in the previous case (Sec. \ref{sec:appdiag1}). To do this we imagine that there is a `ghost right mover'  at $j = -\frac{1}{2}$ at $t=0$ which scatters with all solitons it meets except the first one (the left mover at $j=\frac{1}{2}$). 
		The activation index then goes from $0$ to $1$ at the position where this ghost soliton is found at time $t$. The construction of the tensors is then similar to the previous paragraph, except around that position.
		
		We list all non-zero components. Again, we have
		\begin{eqnarray*}
			&\qquad &	\mathcal{A}[j]_{\sigma, (0,0,0)}^{\sigma, (0,0,0)}  =  	\mathcal{A}[j]_{\sigma, (2,0,0)}^{\sigma, (2,0,0)}   = 1 , \qquad \sigma = True, False . \\
		\end{eqnarray*}
		%Similarly, for $j < -2t$, they are
		%\begin{eqnarray*}
		%	&\qquad &	\mathcal{A}[j]_{\sigma, (2,0,0)}^{\sigma, (2,0,0)}  =  1 , \qquad \sigma = True, False . \\
		%\end{eqnarray*}
		Also, as in the previous paragraph, we have the following componentns when the activation index goes from 1 to 2 (i.e. at the position of the outgoing left mover coming from the origin):
		\begin{eqnarray*}
			j = 2p + \frac{1}{2}, \; p \in \mathbb{Z}: &\qquad &	\mathcal{A}[j]^{True, (2,0,0)}_{True, (1, j,2t - \frac{1}{2})}  = 1  \\
			j = 2p+1, \; p \in \mathbb{Z}: &\qquad &	\mathcal{A}[j]^{True, (2,0,0)}_{True, (1, j-\frac{1}{2},2t - \frac{1}{2})}  = 1  \\
			j = 2p, \; p \in \mathbb{Z}: &\qquad &		\mathcal{A}[j]^{True, (2,0,0)}_{True, (1, j-\frac{3}{2},2t - \frac{1}{2})}  = 1 .
		\end{eqnarray*}
		Inside the region enclosed by the left and right solitons coming from the origin, the activation index is $1$, and we have the following non-zero components:
		\begin{eqnarray*}
			j = 2p - \frac{1}{2}, \; p \in \mathbb{Z} , \; j < j_r -2  : &\qquad &   \mathcal{A}[j]_{True, (1,j_l-2,j_r)}^{True, (1, j_l, j_r)}  = 1  \\
			j = 2p + \frac{1}{2}, \; p \in \mathbb{Z} ,  \;  j  < j_r - 3  : &\qquad &	\mathcal{A}[j]_{True, (1,j_l,j_r-2)}^{True, (1, j_l, j_r)}  = 1  \\
			j = 2p + \frac{1}{2}, \; p \in \mathbb{Z}   : &\qquad &	\mathcal{A}[j]_{True, (1,j_l,j + \frac{3}{2} )}^{True, (1, j_l, j+3)}  =  	\mathcal{A}[j]_{True, (1,j_l,j + \frac{1}{2} )}^{True, (1, j_l, j+1)}  = 1  \\
			%	j = 2p + \frac{1}{2}, \; p \in \mathbb{Z}    : &\qquad &	\mathcal{A}[j]_{True, (1,j_l,j + \frac{1}{2} )}^{True, (1, j_l, j+1)}  = 1  \\
			j  =2p, \; p \in   \mathbb{Z},   \; j < j_r-\frac{7}{2}   : &\qquad &	\mathcal{A}[j]_{True, (1,j_l-2,j_r-2)}^{True, (1, j_l, j_r)}  = 1  \\
			j  =2p+1, \; p \in   \mathbb{Z},   \; j < j_r-\frac{5}{2}   : &\qquad &	\mathcal{A}[j]_{True, (1,j_l-2,j_r-2)}^{True, (1, j_l, j_r)}  = 1  \\
			\forall \,	j  : &\qquad &	\mathcal{A}[j]_{False, (1,j_l,j_r)}^{False, (1, j_l, j_r)}  = 1  .
		\end{eqnarray*}
		and finally, at the position of the ghost right mover, the activation index goes from $1$ to $0$, and the corresponding non-zero components are
		\begin{eqnarray*}
			\forall j :&\qquad &	\mathcal{A}[j]_{False, (0,0,0)}^{False, (1, -2t + \frac{1}{2}, j)}  = 1  .
		\end{eqnarray*}

		\subsection{Third term: $M U^{-t} \left| 1\check{1}0 \right> \left<1 \check{1} 0 \right| U^t M^\dagger$}	
		
		This term is obtained straightforwardly from the previous one (Sec. \ref{sec:appdiag2}) by reflection symmetry $j \rightarrow -j$.

		\subsection{Fourth term: $M U^{-t} \left| 1\check{1}1 \right> \left<1 \check{1} 1 \right| U^t M^\dagger$}	
		\label{sec:appdiag4}
		This is again a minor variation of the first case (Sec. \ref{sec:appdiag1}). We need to make sure that there is a right mover at position $j= - \frac{1}{2}$ and a left mover at $j = \frac{1}{2}$, at time $t=0$. But notice that, since these two solitons will automatically scatter at time $t=1$, this is exactly equivalent to checking that there is a scattering pair at $j=0$ at time $t=1$. So this fourth term is simply related to the first one (Sec. \ref{sec:appdiag1}) by a time shift. The non-zero components are thus exactly the ones of Sec. \ref{sec:appdiag1}, where one makes the replacement $t \rightarrow  t-1$.

		\newpage
		\section{Details on the non-diagonal case}
		\label{sec:app:nondiag}
		
		In the main text, we explained that the operator $ \left| \check{1} \right> \left< \check{0} \right|$ (in the computational basis, and with a `check' indicating the qubit at $j=0$) acting at position $j=0$ must be decomposed as a sum of nine terms that all remain simple upon conjugation by $M$,
		\begin{eqnarray}
		\label{eq:app:nondiag}
		\nonumber	 \left| \check{1} \right> \left< \check{0} \right| \, &= &\, \left| 0\check{1}0 \right>\left< 0\check{0}0 \right|  \, + \,  \left| 0\check{1}10 \right>\left< 0\check{0}10 \right| 
		\, + \,  \left| 0\check{1}11 \right>\left< 0\check{0}11 \right| \\ 
		\nonumber	  && +  \left| 01\check{1}0 \right>\left< 01\check{0}0 \right| \, + \,   \left| 11\check{1}0 \right>\left< 11\check{0}0 \right|  \, + \, \left| 01\check{1}10 \right>\left< 01\check{0}10 \right|  \\
		&& +\left| 01\check{1}11 \right>\left< 01\check{0}11 \right|  \, + \, \left| 11\check{1}10 \right>\left< 11\check{0}10 \right| + \, \left| 11\check{1}11 \right>\left< 11\check{0}11 \right| .
		\end{eqnarray}
		We now explain in detail why each of these nine terms can be written as an MPO with bond dimension growing at most as $\mathcal{O}(t^2)$.

		The general idea is the same for all nine terms. One observes that each term is, upon conjugation by $M$, a sum of the form $\sum \left| s \right> \left< s' \right|$ of equally weighted soliton configurations $s$ and $s'$, where $s'$ is related to $s$ in a specific way. Basically, for each configuration $s$ contributing to the sum, there is a pair of positions $x_1,x_2$ which play a special role, because they correspond to the positions of solitons coming from the origin at $t=0$. Then, for any given $x_1$ and $x_2$, one can construct a linear map $W_{x_1,x_2}$ such that $W_{x_1,x_2} \left| s \right>= \left| s' \right>$ if $x_1,x_2$ are the correct positions for the configurations $s$, and $W_{x_1,x_2} \left| s \right>= 0$ otherwise. Then each of the nine terms can be written in the form
		\begin{equation}
		\label{eq:wanted}
		\sum \left| s \right> \left< s' \right| =   \sum_{x_1,x_2}  \left( \sum \left| s \right> \left< s \right| \right)  W_{x_1,x_2}^\dagger .
		\end{equation}
		$\sum \left| s \right> \left< s \right|$ is a diagonal operator (not the same for all nine terms), and can be written as an MPO with bond dimension at most $\mathcal{O}(t^2)$ according to the discussion of Sec.~\ref{sec:app:diag}. Then the point is that, although the details of the definition of the operator $W_{x_1,x_2}$ are different for all nine terms in Eq. (\ref{eq:app:nondiag}), $W_{x_1, x_2}$ is always an MPO with finite bond dimension, made of tensors $\mathcal{W}_{x_1,x_2}[j]^{\sigma ',b'}_{\sigma,b}= \mathcal{W}[j]^{\sigma ',(a',b')}_{\sigma,(a,b)}$ which do not explicitly depend on $x_1$ or $x_2$, but where $a$ is the same 'activation index' as in the diagonal case,
		\begin{equation*}
		\begin{tikzpicture}
		\begin{scope}
		\draw (-0.5,0) node[left]{\footnotesize $(a',b')$} -- (0.5,0) node[right]{\footnotesize $(a,b)$};
		\draw (0,-0.5) node[below]{\footnotesize $\sigma$}-- (0,0.5) node[above]{\footnotesize $\sigma'$};
		\filldraw[gray!60] (-0.3,-0.3) rectangle (0.3,0.3);
		\draw (0,0) node{$j$};
		\draw (-3.5,0) node{$ \mathcal{W}[j]^{\sigma',(a',b')}_{\sigma,(a,b)} =$};
		\draw (2.5,0);
		\end{scope}
		\end{tikzpicture}
		\end{equation*}
		It is the activation index $a$ that detects the position of $x_1$ and $x_2$, namely:
		\begin{equation*}
		\begin{tikzpicture}
		\begin{scope}
		\draw (-0.5,0) node[left]{\footnotesize $b'$} -- (0.5,0) node[right]{\footnotesize $b$};
		\draw (0,-0.5) node[below]{\footnotesize $\sigma$}-- (0,0.5) node[above]{\footnotesize $\sigma'$};
		\filldraw[blue!60] (-0.3,-0.3) rectangle (0.3,0.3);
		\draw (0,0) node{$j$};
		\draw (-3.5,0) node{$\mathcal{W}_{x_1,x_2}[j]^{\sigma',b'}_{\sigma,b} = \mathcal{W}[j]^{\sigma',(2,b')}_{\sigma,(2,b)} =$};
		\draw (2.5,0) node{if $j < x_1$,};
		\end{scope}
		\begin{scope}[yshift=-2cm]
		\draw (-0.5,0) node[left]{\footnotesize $b'$} -- (0.5,0) node[right]{\footnotesize $b$};
		\draw (0,-0.5) node[below]{\footnotesize $\sigma$}-- (0,0.5) node[above]{\footnotesize $\sigma'$};
		\filldraw[red!60] (-0.3,-0.3) rectangle (0.3,0.3);
		\draw (0,0) node{$j$};
		\draw (-3.5,0) node{$\mathcal{W}_{x_1,x_2}[j]^{\sigma',b'}_{\sigma,b} = \mathcal{W}[j]^{\sigma',(2,b')}_{\sigma,(1,b)} =$};
		\draw (2.5,0) node{if $j = x_1$,};
		\end{scope}
		\begin{scope}[yshift=-4cm]
		\draw (-0.5,0) node[left]{\footnotesize $b'$} -- (0.5,0) node[right]{\footnotesize $b$};
		\draw (0,-0.5) node[below]{\footnotesize $\sigma$}-- (0,0.5) node[above]{\footnotesize $\sigma'$};
		\filldraw[yellow!60] (-0.3,-0.3) rectangle (0.3,0.3);
		\draw (0,0) node{$j$};
		\draw (-3.8,0) node{$\mathcal{W}_{x_1,x_2}[j]^{\sigma',b'}_{\sigma,b} = \mathcal{W}[j]^{\sigma',(1,b')}_{\sigma,(1,b)} =$};
		\draw (2.5,0) node{if $x_1 < j < x_2$,};
		\end{scope}
		\begin{scope}[yshift=-6cm]
		\draw (-0.5,0) node[left]{\footnotesize $b'$} -- (0.5,0) node[right]{\footnotesize $b$};
		\draw (0,-0.5) node[below]{\footnotesize $\sigma$}-- (0,0.5) node[above]{\footnotesize $\sigma'$};
		\filldraw[green!60] (-0.3,-0.3) rectangle (0.3,0.3);
		\draw (0,0) node{$j$};
		\draw (-3.5,0) node{$\mathcal{W}_{x_1,x_2}[j]^{\sigma',b'}_{\sigma,b} = \mathcal{W}[j]^{\sigma',(1,b')}_{\sigma,(0,b)} =$};
		\draw (2.5,0) node{if $j = x_2$,};
		\end{scope}
		\begin{scope}[yshift=-8cm]
		\draw (-0.5,0) node[left]{\footnotesize $b'$} -- (0.5,0) node[right]{\footnotesize $b$};
		\draw (0,-0.5) node[below]{\footnotesize $\sigma$}-- (0,0.5) node[above]{\footnotesize $\sigma'$};
		\filldraw[blue!60] (-0.3,-0.3) rectangle (0.3,0.3);
		\draw (0,0) node{$j$};
		\draw (-3.5,0) node{$\mathcal{W}_{x_1,x_2}[j]^{\sigma',b'}_{\sigma,b} = \mathcal{W}[j]^{\sigma',(0,b')}_{\sigma,(0,b)} =$};
		\draw (2.5,0) node{if $j > x_2$,};
		\end{scope}
		\end{tikzpicture}
		\end{equation*}
		These tensors are contracted as
		\begin{equation*}
		\begin{tikzpicture}
		\draw (-7,0) node{$W_{x_1,x_2} =$};
		\draw[dashed] (-6,0) -- (-5.3,0);
		\draw[dashed] (6,0) -- (5.3,0);
		\begin{scope}
		\draw (-5.5,0) -- (5.5,0);
		\foreach \x in {-5,-4,...,5} {
			\draw (\x,-0.5) -- (\x,0.5);
		}
		\filldraw[blue!60] (-0.3-5,-0.3) rectangle ++(0.6,0.6);
		\filldraw[blue!60] (-0.3-4,-0.3) rectangle ++(0.6,0.6);
		\filldraw[red!60] (-0.3-3,-0.3) rectangle ++(0.6,0.6);
		\filldraw[yellow!60] (-0.3-2,-0.3) rectangle ++(0.6,0.6);
		\filldraw[yellow!60] (-0.3-1,-0.3) rectangle ++(0.6,0.6);
		\filldraw[yellow!60] (-0.3+0,-0.3) rectangle ++(0.6,0.6);
		\filldraw[yellow!60] (-0.3+1,-0.3) rectangle ++(0.6,0.6);
		\filldraw[green!60] (-0.3+2,-0.3) rectangle ++(0.6,0.6);
		\filldraw[blue!60] (-0.3+3,-0.3) rectangle ++(0.6,0.6);
		\filldraw[blue!60] (-0.3+4,-0.3) rectangle ++(0.6,0.6);
		\filldraw[blue!60] (-0.3+5,-0.3) rectangle ++(0.6,0.6);
		\draw (-3,0) node{$x_1$};
		\draw (2,0) node{$x_2$};
		\end{scope}
		\end{tikzpicture}
		\end{equation*}
		Then, assuming that we already have an MPO $ \mathcal{A} $ for the diagonal operator,
		\begin{equation*}
		\begin{tikzpicture}
		\draw (-7,0) node{$\sum \left| s \right> \left< s \right| =$};
		\draw[dashed] (-6,0) -- (-5.3,0);
		\draw[dashed] (6,0) -- (5.3,0);
		\begin{scope}
		\draw (-5.5,0) -- (5.5,0);
		\foreach \x in {-5,-4,...,5} {
			\draw (\x,-0.5) -- (\x,0.5);
		}
		\filldraw[red] (-0.3-5,-0.3) rectangle ++(0.6,0.6);
		\filldraw[red] (-0.3-4,-0.3) rectangle ++(0.6,0.6);
		\filldraw[red] (-0.3-3,-0.3) rectangle ++(0.6,0.6);
		\filldraw[red] (-0.3-2,-0.3) rectangle ++(0.6,0.6);
		\filldraw[red] (-0.3-1,-0.3) rectangle ++(0.6,0.6);
		\filldraw[red] (-0.3+0,-0.3) rectangle ++(0.6,0.6);
		\filldraw[red] (-0.3+1,-0.3) rectangle ++(0.6,0.6);
		\filldraw[red] (-0.3+2,-0.3) rectangle ++(0.6,0.6);
		\filldraw[red] (-0.3+3,-0.3) rectangle ++(0.6,0.6);
		\filldraw[red] (-0.3+4,-0.3) rectangle ++(0.6,0.6);
		\filldraw[red] (-0.3+5,-0.3) rectangle ++(0.6,0.6);
		\end{scope}
		\end{tikzpicture}
		\end{equation*}
		we can adapt it to build an MPO for the non-diagonal case, by matching the activation index $(a,a') $ of the MPO for $ \mathcal{W} $ and $ \mathcal{A} $. Thus, one defines a new tensor
		\begin{equation*}
		\begin{tikzpicture}
		\begin{scope}
		\begin{scope}
		\draw (-0.5,0) node[left]{\footnotesize $(a',b',j_l',j_r')$} -- (0.5,0) node[right]{\footnotesize $(a,b,j_l,j_r)$};
		\draw (0,-0.5) node[below]{\footnotesize $\sigma$}-- (0,0.5) node[above]{\footnotesize $\sigma'$};
		\filldraw[orange] (-0.3,-0.3) rectangle (0.3,0.3);
		\draw (0,0) node{$j$};
		\end{scope}
		\draw (2.5,0) node{$\equiv$};
		\begin{scope}[xshift=5cm,yshift=-0.5cm]
		\draw (-0.5,0) node[left]{\footnotesize $(a',b')$} -- (0.5,0) node[right]{\footnotesize $(a,b)$};
		\draw (-0.5,0.8) node[left]{\footnotesize $(a',j_l',j_r')$} -- (0.5,0.8) node[right]{\footnotesize $(a,j_l,j_r)$};
		\draw (0,-0.5) node[below]{\footnotesize $\sigma$}-- (0,1.3) node[above]{\footnotesize $\sigma'$};
		\filldraw[gray!60] (-0.3,-0.3) rectangle (0.3,0.3);
		\filldraw[red] (-0.3,0.5) rectangle (0.3,1.1);
		\draw (0,0) node{$j$};
		\draw (0,0.8) node{$j$};
		\end{scope}
		\end{scope}
		\end{tikzpicture}
		\end{equation*}
		such that the contraction
		\begin{equation*}
		\begin{tikzpicture}
		%	\draw (-7,0) node{$\sum \left| s \right> \left< s \right| =$};
		\draw[dashed] (-6,0) -- (-5.3,0);
		\draw[dashed] (6,0) -- (5.3,0);
		\begin{scope}
		\draw (-5.5,0) -- (5.5,0);
		\foreach \x in {-5,-4,...,5} {
			\draw (\x,-0.5) -- (\x,0.5);
		}
		\filldraw[orange] (-0.3-5,-0.3) rectangle ++(0.6,0.6);
		\filldraw[orange] (-0.3-4,-0.3) rectangle ++(0.6,0.6);
		\filldraw[orange] (-0.3-3,-0.3) rectangle ++(0.6,0.6);
		\filldraw[orange] (-0.3-2,-0.3) rectangle ++(0.6,0.6);
		\filldraw[orange] (-0.3-1,-0.3) rectangle ++(0.6,0.6);
		\filldraw[orange] (-0.3+0,-0.3) rectangle ++(0.6,0.6);
		\filldraw[orange] (-0.3+1,-0.3) rectangle ++(0.6,0.6);
		\filldraw[orange] (-0.3+2,-0.3) rectangle ++(0.6,0.6);
		\filldraw[orange] (-0.3+3,-0.3) rectangle ++(0.6,0.6);
		\filldraw[orange] (-0.3+4,-0.3) rectangle ++(0.6,0.6);
		\filldraw[orange] (-0.3+5,-0.3) rectangle ++(0.6,0.6);
		\end{scope}
		\end{tikzpicture}
		\end{equation*}
		is exactly Eq. (\ref{eq:wanted}). So it is an MPO for the specific term we are looking at in the sum (\ref{eq:app:nondiag}). The crucial point is that, because the tensors $\mathcal{W}[j]$ have finite bond dimension (i.e. the index $p$ lives in some finite set, independent of time $t$), the scaling of the total bond dimension with $t$ remains $\mathcal{O}(t^2)$ as claimed in the main text.

		\subsection{First term in Eq. (\ref{eq:app:nondiag}): $M U^{-t} \left| 0\check{1}0 \right> \left<0 \check{0} 0 \right| U^t M^\dagger$}	
		\label{sec:app:case1}

		\begin{figure}[h]
			\begin{tikzpicture}
			\draw (0,0) node{\includegraphics[width=0.35\textwidth]{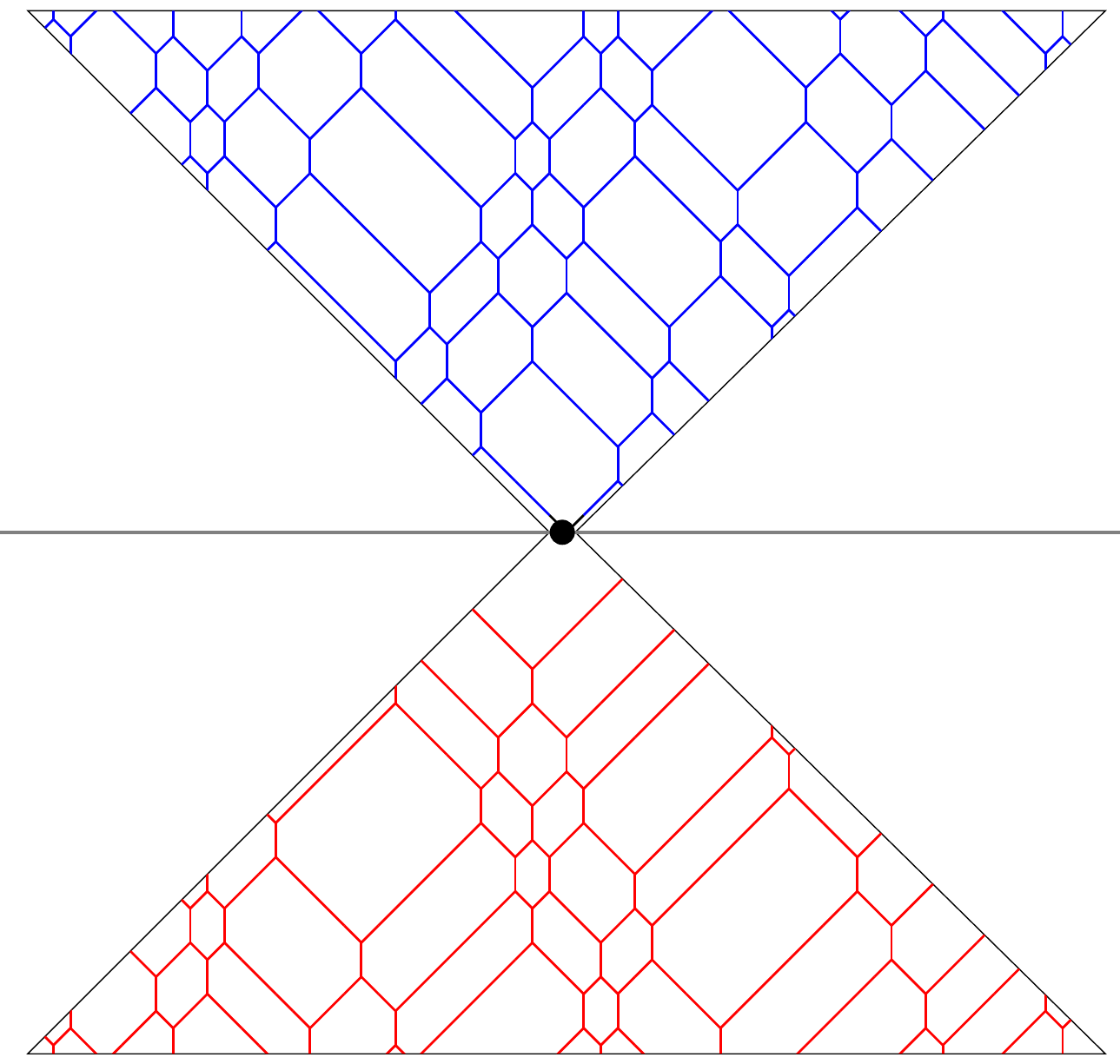}};
			\draw (7,0.4) node{\includegraphics[width=0.4\textwidth]{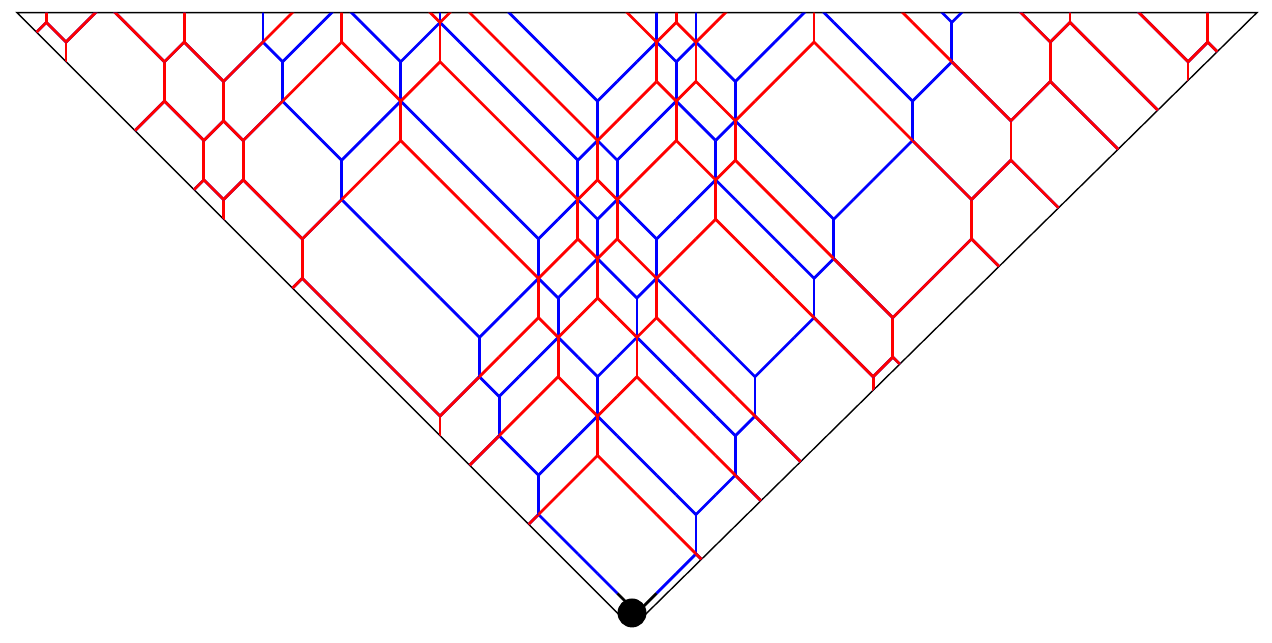}};
			\draw[<-] (4.9,2.2) -- ++(0,0.4) node[above]{$x_1$};
			\draw[<-] (8.83,2.2) -- ++(0,0.4) node[above]{$x_2$};
			\end{tikzpicture}
			\caption{A typical soliton configuration contributing to $M U^{-t} \left| 0\check{1}0 \right> \left<0 \check{0} 0 \right| U^t M^\dagger$. The key observation is that, after folding, the blue configuration and the red one are related by a time-shift of one time unit inside the region enclosed by the left and right moving soliton that came from the origin.}
			\label{fig:app:nondiag1}
		\end{figure}
		The key observation (see Fig. \ref{fig:app:nondiag1}) is that 
		\begin{equation}
		M U^{-t} \left| 0\check{1}0 \right> \left<0 \check{0} 0 \right| U^t M^\dagger =  \sum_{x_1,x_2} (M U^{-t} \ket{ 0\check{1}0 } \left<0 \check{1} 0 \right| U^t M^\dagger)   W_{x_1,x_2}^\dagger
		\end{equation}
		where $(M U^{-t} \left| 0\check{1}0 \right> \left<0 \check{1} 0 \right| U^t M^\dagger)$ is the diagonal operator of section \ref{sec:appdiag1} ---which can be written as an MPO with bond dimension $\mathcal{O}(t^2)$---, and $W_{x_1,x_2}$ is the operator which
		\begin{itemize}
			\item erases the right mover at position $x_2$ and the left mover at position $x_1$
			\item acts as the identity outside the interval $(x_1,x_2)$
			\item acts as the evolution operator (applying a time shift of one time unit) inside the interval $(x_1,x_2)$.
		\end{itemize}

		We start by writing the evolution operator in the soliton basis, $M U M^\dagger$, as an MPO with bond dimension $3$. The building block of that MPO is the tensor $\mathcal{U}[j]_{\sigma, b}^{\sigma', b'}$ with $b,b'=0,1,2$, with the following non-zero components:
		\begin{eqnarray*}
			j= 2p + \frac{1}{2}, \; p \in\mathbb{Z}  \quad ({\rm left\; moving})  : &\quad & \mathcal{U}[j]_{False, 0}^{False, 0} = \mathcal{U}[j]_{False, 2}^{False, 2} =  \mathcal{U}[j]_{True, 0}^{False, 1} = \mathcal{U}[j]_{False, 1}^{True, 0}= 1 \\
			j= 2p - \frac{1}{2}, \; p \in\mathbb{Z}  \quad ({\rm right\; moving})  : &\quad & \mathcal{U}[j]_{False, 0}^{False, 0} = \mathcal{U}[j]_{False, 1}^{False, 1}= \mathcal{U}[j]_{True, 2}^{False, 0} = \mathcal{U}[j]_{False, 0}^{True, 2}  = 1 \\
			j \in\mathbb{Z} \quad ({\rm scattering \; pair}) : &\quad & \mathcal{U}[j]_{False, 0}^{False, 0} = \mathcal{U}[j]_{False, 1}^{False, 1}  =   \mathcal{U}[j]_{False, 2}^{False, 2}=   \mathcal{U}[j]_{True, 2}^{False, 1}  =   \mathcal{U}[j]_{False, 1}^{True, 2}=  1 . %\\
			%	j= 2p+1, \; p \in\mathbb{Z} \quad ({\rm fusing\; pair}): &\quad & \mathcal{U}[j]_{False, 0}^{False, 0} = \mathcal{U}[j]_{False, 1}^{False, 1}  =   \mathcal{U}[j]_{False, 2}^{False, 2}=   \mathcal{U}[j]_{True, 2}^{False, 1}  =   \mathcal{U}[j]_{False, 1}^{True, 2}=  1
		\end{eqnarray*}
		The idea here is that the auxiliary state $0$ indicates the absence of a soliton, $1$ stands for a left mover, and $2$ stands for a right mover. Then the non-zero components are chose in order to implement the basic moves of solitons. \vspace{0.4cm}

		Then we define the tensors that allow to write $W_{x_1,x_2}$ as an MPO as follows. The components are written as $\mathcal{W}[j]_{\sigma,(a,b)}^{\sigma', (a',b')} $ (see the introduction to Sec. \ref{sec:app:nondiag} above). On the left of $x_1$, the activation index is $2$, and $W_{x_1,x_2}$ acts as the identity. This is implemented by the non-zero components
		$$
		\mathcal{W}[j]_{\sigma,(2,0)}^{\sigma, (2,0)} =  1.
		$$
		At position $x_1$, the activation index goes from $2$ to $1$, and the non-zero components are chosen as
		$$
		\mathcal{W}[j]_{True,(1,b)}^{\sigma', (2,0)} = \mathcal{U}[j]_{True, b}^{\sigma',1} .
		$$
		Between $x_1$ and $x_2$, the activation index is always $1$, and the non-zero components are chose in order for $W_{x_1,x_2}$ to act as the evolution operator,
		$$
		\mathcal{W}[j]_{\sigma,(1,b)}^{\sigma', (1,b')} = \mathcal{U}[j]_{\sigma, b}^{\sigma', b'} .
		$$
		At position $x_2$, the activation index goes from $1$ to $0$, and the non-zero components are
		$$
		\mathcal{W}[j]_{True,(0,0)}^{\sigma', (1,b')} = \mathcal{U}[j]_{True, 2}^{\sigma',b'} .
		$$
		Finally, on the right of $x_2$, the activation index is $0$. $W_{x_1,x_2}$ acts again as the identity, and this is implemented by the non-zero components
		$$
		\mathcal{W}[j]_{\sigma,(0,0)}^{\sigma, (0,0)} =  1.
		$$

		\subsection{Second term in Eq. (\ref{eq:app:nondiag}): $M U^{-t} \left| 0\check{1}10 \right> \left<0 \check{0} 10 \right| U^t M^\dagger$}	
		\label{sec:app:case2}
		
		\begin{figure}[h]
			\begin{tikzpicture}
			\draw (0,0) node{\includegraphics[width=0.35\textwidth]{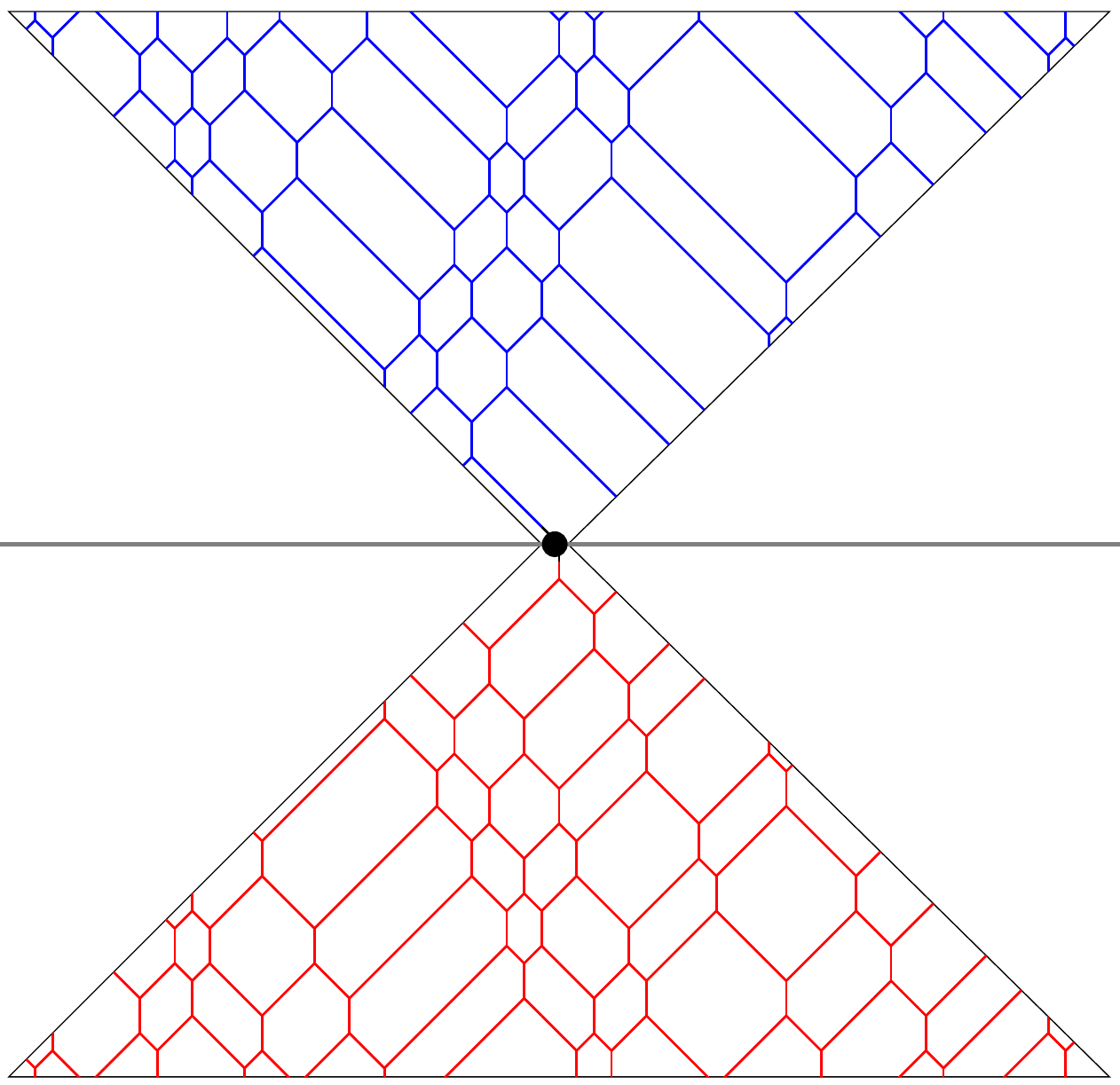}};
			\draw (7,0.4) node{\includegraphics[width=0.4\textwidth]{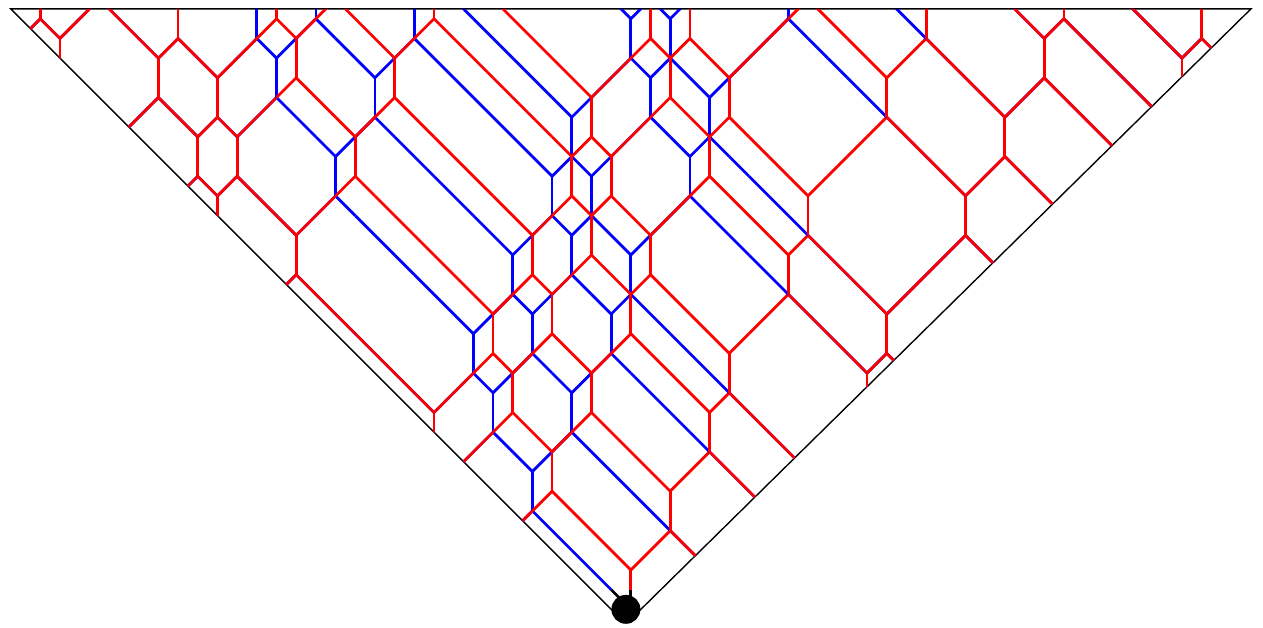}};
			\draw[<-] (4.87,2.2) -- ++(0,0.4) node[above]{$x_1$};
			\draw[<-] (8.69,2.2) -- ++(0,0.4) node[above]{$x_2$};
			%	\draw[<-] (8.12,2.2) -- ++(0,0.4) node[above]{$x_2'$};
			\end{tikzpicture}
			\caption{A typical soliton configuration contributing to $M U^{-t} \left| 0\check{1}10 \right> \left<0 \check{0} 10 \right| U^t M^\dagger$. After folding, the blue configuration is obtained from the red one by applying a backward time-shift of one half unit time and a translation of one site to the right inside the interval $(x_1,x_2)$ enclosed by the left and right moving soliton that came from the origin.}
			\label{fig:app:case2}
		\end{figure}
		
		We write the second term as
		\begin{equation}
		\label{eq:app:case2}
		M U^{-t} \left| 0\check{1}10 \right> \left<0 \check{0} 10 \right| U^t M^\dagger =  \sum_{x_1,x_2}  (M U^{-t} \left| 0\check{1} 1  \right> \left< 0\check{1} 1 \right| U^t M^\dagger ) W_{x_1,x_2}^\dagger
		\end{equation}
		where $W_{x_1, x_2}$ is an operator which (see Fig. \ref{fig:app:case2})
		\begin{itemize}
			%	\item  finds the position $x_2'$ of the first right mover on the left of $x_2$
			%		and checks that $x_2'$ is sufficiently far from $x_2$ (in order to make sure it did not come from position $j = -\frac{1}{2}$ at $t=0$, so that the bit at $j=-1$ was indeed in the state '$0$')
			\item creates a right mover at position $x_2$
			\item applies a time-shift (by a half-time step, backwards) and a translation (by one site to the right) inside the interval $(x_1,x_2)$
			\item acts as the identity outside the interval $(x_1,x_2)$.
		\end{itemize}
		With that definition, $W_{x_1,x_2}$ may produce soliton configurations which do not correspond to any qubit configuration. For instance, it can produce configurations with two right movers at $j$ and $j+2$ (and no left mover at $j+1$), which does not correspond to any qubit configuration. However, when one conjugates the resulting MPO by $M^\dagger$ in order to get $M^\dagger   \sum_{x_1,x_2}  (M U^{-t} \left| 0\check{1} 1 \right> \left<  0\check{1} 1  \right| U^t M^\dagger ) W_{x_1,x_2}^\dagger M$, all such non-admissible soliton configurations are projected out. It turns out that the configurations that remain with a non-zero amplitude are exactly the ones with no right mover at $j=\frac{1}{2}$ at $t=0$, ensuring that the central qubit configuration at $t=0$ is indeed '$0\check{1}10$', and not '$0\check{1}11$'. This is exactly what is needed in order for Eq. (\ref{eq:app:case2}) to hold.

		$W_{x_1,x_2}$ can be written as an MPO as follows. First, we write the MPO that implements the time-shift and the translation. We decompose the two operations. The MPO that implements the backward time-evolution on a half time unit is written with tensors $\mathcal{U}^{-\frac{1}{2}}[j]^{\sigma', b'}_{\sigma, b}$ with the following components:
		\begin{eqnarray*}
			j= 2p + \frac{1}{2}, \; p \in\mathbb{Z}  \quad ({\rm left\; moving})  : &\quad &  \mathcal{U}^{-\frac{1}{2}}[j]_{False, 0}^{False, 0} =  \mathcal{U}^{-\frac{1}{2}}[j]_{True, 1}^{False, 0} = \mathcal{U}^{-\frac{1}{2}}[j]_{False, 2}^{True, 0}= 1 \\
			j= 2p - \frac{1}{2}, \; p \in\mathbb{Z}  \quad ({\rm right\; moving})  : &\quad & \mathcal{U}^{-\frac{1}{2}}[j]_{False, 0}^{False, 0} =  \mathcal{U}^{-\frac{1}{2}}[j]_{True, 0}^{False, 2} = \mathcal{U}^{-\frac{1}{2}}[j]_{False, 0}^{True, 1}  = 1 \\
			j =2p, \; p \in\mathbb{Z}  \in\mathbb{Z} \quad ({\rm splitting \; pair}) : &\quad &     \mathcal{U}^{-\frac{1}{2}}[j]_{True, 0}^{True, 0}  =  \mathcal{U}^{-\frac{1}{2}}[j]_{False, 0}^{False, 0}     \\	
			j= 2p+1, \; p \in\mathbb{Z} \quad ({\rm fusing\; pair}): &\quad &    \mathcal{U}^{-\frac{1}{2}}[j]_{True, 1}^{False, 2}=      \mathcal{U}^{-\frac{1}{2}}[j]_{False, 2}^{True, 1}=    \mathcal{U}^{-\frac{1}{2}}[j]_{False, 0}^{False, 0} =   \mathcal{U}^{-\frac{1}{2}}[j]_{False, 1}^{False, 1}=   \mathcal{U}^{-\frac{1}{2}}[j]_{False, 2}^{False, 2} =1 .
		\end{eqnarray*}
		The translation by one site to the right is written as an MPO with tensors $\mathcal{T}[j]^{\sigma', b'}_{\sigma, b}$ that have non-zero components
		\begin{equation}
		\mathcal{T}[j]^{False, 0}_{False, 0} = \mathcal{T}[j]^{False, 0}_{True, 1}= \mathcal{T}[j]^{False, 1}_{False, 2}  =  \mathcal{T}[j]^{True, 2}_{False, 0} =\mathcal{T}[j]^{True, 2}_{True, 1}    .
		\end{equation}
		The composition of the two operations can be written as an MPO with tensors $\mathcal{T} \cdot \mathcal{U}^{-\frac{1}{2}}[j]$ defined as (for notational convenience we group the indices $b = (b_1, b_2)$)
		$$
		\mathcal{T} \cdot \mathcal{U}^{-\frac{1}{2}}[j]_{\sigma, b}^{\sigma', b'} = \mathcal{T} \cdot \mathcal{U}^{-\frac{1}{2}}[j]_{\sigma, (b_1,b_2)}^{\sigma', (b_1',b_2')} \equiv   \sum_{\sigma''}  \mathcal{T}[j]^{\sigma', b_1'}_{\sigma'', b_1} \mathcal{U}^{-\frac{1}{2}}[j]^{\sigma'', b_2'}_{\sigma, b_2} .
		$$
		This then gives an MPO with finite bond dimension (the bond dimension is 9 here, since $b_1$ and $b_2$ both go from $0$ to $2$).

		Next, we define new tensors $\mathcal{W} [j]^{\sigma', (a', b')}_{\sigma, (a,b)}$ (see the introduction of Sec. \ref{sec:app:nondiag} in this Supplementary Material) with the following non-zero components. For sites on the left of $x_1$, the activation index $a$ is two, and the corresponding non-zero components are
		\begin{eqnarray*}
			&& \mathcal{W} [j]^{\sigma, (2,0)}_{\sigma, (2,0)}  = 1 
			%	&& \mathcal{W}_{x_2<.} [j]^{True, (0,0)}_{True, (0,0)} = \mathcal{W}_{x_2<.} [j]^{False, (0,0)}_{False, (0,0)} = 1
		\end{eqnarray*}
		which ensure that the MPO acts as the identity in that region. Similarly, on the right of $x_2$, the activation index is zero, and the non-zero components are
		\begin{eqnarray*}
			&& \mathcal{W} [j]^{\sigma, (0,0)}_{\sigma, (0,0)} =   1  .
			%	&& \mathcal{W}_{x_2<.} [j]^{True, (0,0)}_{True, (0,0)} = \mathcal{W}_{x_2<.} [j]^{False, (0,0)}_{False, (0,0)} = 1
		\end{eqnarray*}
		At position $j=x_1$, the activation index changes from $1$ to $2$. Here there are two cases we need to distinguish. If the left mover coming from the origin is in a fusing pair at time $t$ (i.e. if $x_1 \in 2 \mathbb{Z}+1$) then we need to be careful because evolving the pair backwards would create a right mover at position $x_1 - \frac{1}{2}$, outside the interval $[x_1,x_2]$. However this is easily taken care of by appropriately fixing the index $b' = (b_1',b_2')$ to $(1,2)$ in the operator $\mathcal{T} \cdot \mathcal{U}^{-\frac{1}{2}}$ at this position. If the left mover coming from the origin is not in a fusing pair, then the index $b' = (b_1', b_2')$ simply needs to be fixed to $(0,0)$. The corresponding non-zero components are
		\begin{eqnarray*}
			j \in 2\mathbb{Z}+1 \quad ({\rm fusing \; pair}):  &\quad & \mathcal{W} [j]^{\sigma, (2,0)}_{True, (1,b)} = \mathcal{T}\cdot \mathcal{U}^{\frac{1}{2}}[j]_{True, b}^{\sigma, (1,2)} , \\
			j=  \notin 2\mathbb{Z}+1 \quad ({\rm not \; a \; fusing \; pair}):  &\quad & \mathcal{W} [j]^{\sigma, (2,0)}_{True, (1,b)} = \mathcal{T}\cdot \mathcal{U}^{\frac{1}{2}}[j]_{True, b}^{\sigma, (0,0)}  .
		\end{eqnarray*}
		Inside the interval $(x_1,x_2)$, the activation index is $1$, and one implements the time-shift and the translation with the non-zero components
		\begin{eqnarray*}
			\mathcal{W} [j]^{\sigma', (1,b')}_{\sigma, (1,b)} = \mathcal{T}\cdot \mathcal{U}^{\frac{1}{2}}[j]_{\sigma, b}^{\sigma', b'} .
		\end{eqnarray*}
		At $x_2$, the activation index switches form $0$ to $1$, and one need to create an additional right mover. This is done with the non-zero components
		\begin{eqnarray*}
			& & \mathcal{W} [j]^{True, (1,b')}_{False, (0,0)} =1 
		\end{eqnarray*}
		for all $j$ and $b'$.

		\subsection{Third term in Eq. (\ref{eq:app:nondiag}): $M U^{-t} \left| 0\check{1}11 \right> \left<0 \check{0} 11 \right| U^t M^\dagger$}	
		
		\label{sec:app:case3}

		\begin{figure}[h]
			\begin{tikzpicture}
			\draw (0,0) node{\includegraphics[width=0.35\textwidth]{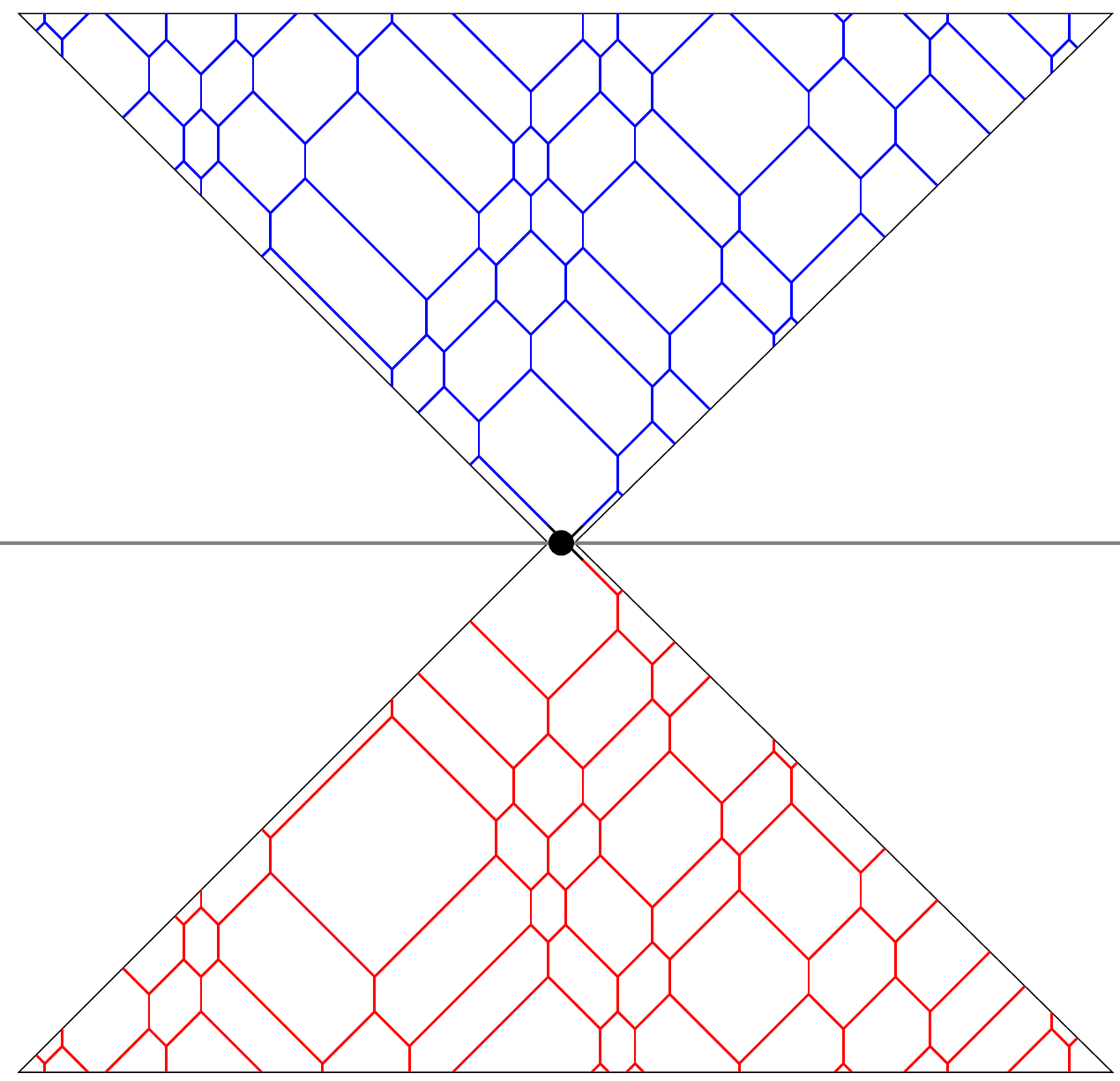}};
			\draw (7,0.4) node{\includegraphics[width=0.4\textwidth]{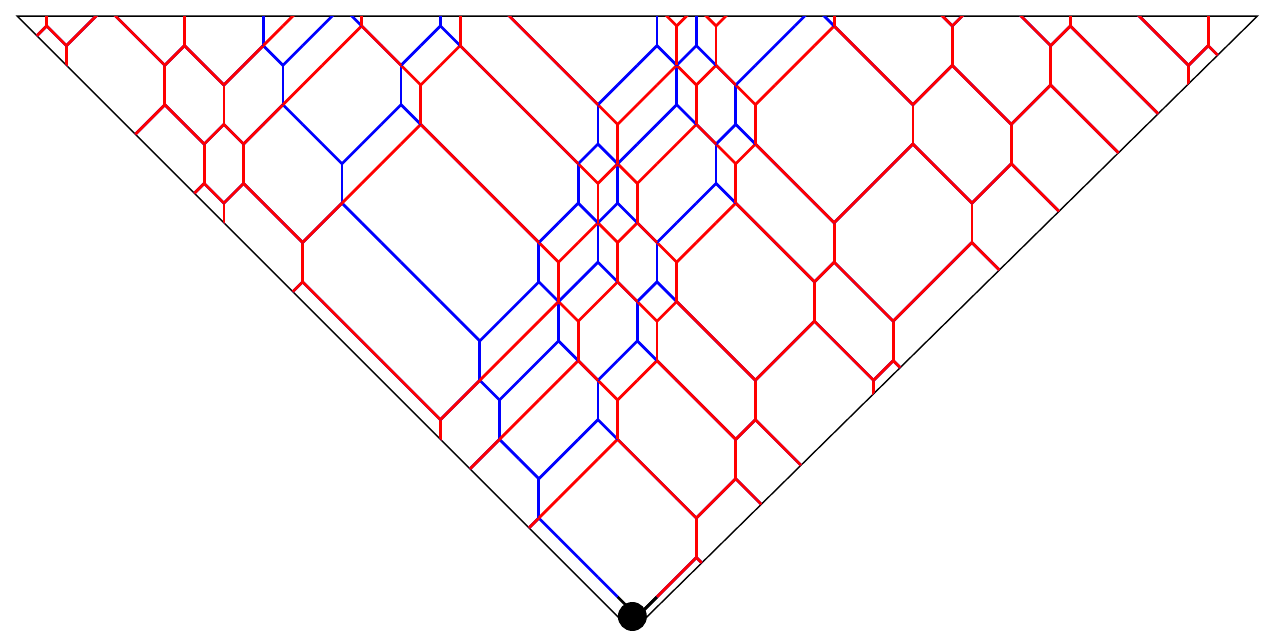}};
			\draw[<-] (4.9,2.2) -- ++(0,0.4) node[above]{$x_1$};
			\draw[<-] (8.83,2.2) -- ++(0,0.4) node[above]{$x_2$};
			%	\draw[<-] (8.12,2.2) -- ++(0,0.4) node[above]{$x_2'$};
			\end{tikzpicture}
			\caption{A typical soliton configuration contributing to $M U^{-t} \left| 0\check{1}11 \right> \left<0 \check{0} 11 \right| U^t M^\dagger$. The key observation is that, after folding, the red configuration is obtained from the blue one by applying a time-shift of one half unit time and a translation of one site to the right inside the interval $(x_1,x_2)$ enclosed by the left and right moving soliton that came from the origin.}
			\label{fig:app:case3}
		\end{figure}
		
		%First, for convenience, we translate the whole system by one site, so that the operator we are interested in is $M U^{-t} \left| 01\check{1}1 \right> \left<0 0\check{1} 1 \right| U^t M^\dagger$, where the 'check' indicates the bit at the origin $j=0$ (this obviously does not make any difference, but it saves us from having to implement this shift explicitly in all formulas for the components). Next, we write 
		This time we write
		\begin{equation}
		\label{eq:app:case3}
		M U^{-t} \left| 0\check{1}11 \right> \left<0 \check{0} 11 \right| U^t M^\dagger =  \sum_{x_1,x_2}  (M U^{-t} \left| \check{1} 1 1 \right> \left< \check{1} 1 1 \right| U^t M^\dagger ) W_{x_1,x_2}^\dagger
		\end{equation}
		where $W_{x_1, x_2}$ is an operator which (see Fig. \ref{fig:app:case3})
		\begin{itemize}
			%	\item  finds the position $x_2'$ of the first right mover on the left of $x_2$
			%		and checks that $x_2'$ is sufficiently far from $x_2$ (in order to make sure it did not come from position $j = -\frac{1}{2}$ at $t=0$, so that the bit at $j=-1$ was indeed in the state '$0$')
			\item erases the soliton at position $x_1$
			\item applies a time-shift (by a half-time step) and a translation (by one site to the right) inside the interval $(x_1,x_2)$
			\item acts as the identity outside the interval $(x_1,x_2)$.
		\end{itemize}
		Again, there is an important subtlety: with that definition $W_{x_1,x_2}$, may produce soliton configurations which do not correspond to any spin configuration. For instance, it can produce configurations with two neighboring right movers (and no left mover between them), which does not correspond to any spin configuration. However, when one conjugates the resulting MPO by $M^\dagger$ in order to get $M^\dagger   \sum_{x_1,x_2}  (M U^{-t} \left| \check{1} 1 1 \right> \left< \check{1} 1 1 \right| U^t M^\dagger ) W_{x_1,x_2}^\dagger M$, all such non-admissible soliton configurations are projected out. It turns out that the configurations that remain with a non-zero amplitude are exactly the ones with no right mover initially on the left of the pair of solitons coming from the origin at $t=0$, or in other words, ensuring that the central qubit configuration at $t=0$ is indeed '$0\check{1}11$', and not '$1\check{1}11$'. This is exactly what is needed in order for Eq. (\ref{eq:app:case3}) to hold.

		$W_{x_1,x_2}$ can be written as an MPO as follows. First, we write the non-zero components of the MPO that implements the time-shift and the translation. We decompose the two operations. The MPO that implements the time-evolution on a half time unit is written with tensors $\mathcal{U}^{\frac{1}{2}}[j]^{\sigma', b'}_{\sigma, b}$ with the following components:
		\begin{eqnarray*}
			j= 2p + \frac{1}{2}, \; p \in\mathbb{Z}  \quad ({\rm left\; moving})  : &\quad &  \mathcal{U}^{\frac{1}{2}}[j]_{False, 0}^{False, 0} =  \mathcal{U}^{\frac{1}{2}}[j]_{True, 0}^{False, 1} = \mathcal{U}^{\frac{1}{2}}[j]_{False, 0}^{True, 2}= 1 \\
			j= 2p - \frac{1}{2}, \; p \in\mathbb{Z}  \quad ({\rm right\; moving})  : &\quad & \mathcal{U}^{\frac{1}{2}}[j]_{False, 0}^{False, 0} =  \mathcal{U}^{\frac{1}{2}}[j]_{True, 2}^{False, 0} = \mathcal{U}^{\frac{1}{2}}[j]_{False, 1}^{True, 0}  = 1 \\
			j =2p, \; p \in\mathbb{Z}  \in\mathbb{Z} \quad ({\rm splitting \; pair}) : &\quad &      \mathcal{U}^{\frac{1}{2}}[j]_{True, 2}^{False, 1}=      \mathcal{U}^{\frac{1}{2}}[j]_{False, 1}^{True, 2}=    \mathcal{U}^{\frac{1}{2}}[j]_{False, 0}^{False, 0} =   \mathcal{U}^{\frac{1}{2}}[j]_{False, 1}^{False, 1}=   \mathcal{U}^{\frac{1}{2}}[j]_{False, 2}^{False, 2} =1  \\
			j= 2p+1, \; p \in\mathbb{Z} \quad ({\rm fusing\; pair}): &\quad &   \mathcal{U}^{\frac{1}{2}}[j]_{True, 0}^{True, 0}  =  \mathcal{U}^{\frac{1}{2}}[j]_{False, 0}^{False, 0}   .
		\end{eqnarray*}
		The translation by one site to the right is written as an MPO with tensors $\mathcal{T}[j]^{\sigma', b'}_{\sigma, b}$ that have non-zero components
		\begin{equation}
		\mathcal{T}[j]^{False, 0}_{False, 0} = \mathcal{T}[j]^{False, 0}_{True, 1}= \mathcal{T}[j]^{False, 1}_{False, 2}  = \mathcal{T}[j]^{True, 2}_{False,0}   = \mathcal{T}[j]^{True, 2}_{True, 1} .
		\end{equation}
		The composition of the two operations can be written as an MPO with tensors $\mathcal{T} \cdot \mathcal{U}^{\frac{1}{2}}[j]$ defined as (for notational convenience we group the indices $b = (b_1, b_2)$)
		$$
		\mathcal{T} \cdot \mathcal{U}^{\frac{1}{2}}[j]_{\sigma, b}^{\sigma', b'} = \mathcal{T} \cdot \mathcal{U}^{\frac{1}{2}}[j]_{\sigma, (b_1,b_2)}^{\sigma', (b_1',b_2')} \equiv   \sum_{\sigma''}  \mathcal{T}[j]^{\sigma', b_1'}_{\sigma'', b_1} \mathcal{U}^{\frac{1}{2}}[j]^{\sigma'', b_2'}_{\sigma, b_2} .
		$$
		This then gives an MPO with finite bond dimension (the bond dimension is 9 here, since $b_1$ and $b_2$ both go from $0$ to $2$).

		Next, we define new tensors $\mathcal{W} [j]^{\sigma', (a',b')}_{\sigma, (a,b)}$ (see the introduction of Sec. \ref{sec:app:nondiag} in this Supplementary Material) with the following non-zero components. For sites on the left of $x_1$, the activation index is $2$ and the corresponding non-zero components are
		\begin{eqnarray*}
			&& \mathcal{W} [j]^{\sigma, (2,0)}_{\sigma, (2,0)} = 1 ,
		\end{eqnarray*}
		which ensure that the MPO acts as the identity in that region. At $x_1$ (i.e. where the activation index goes from $2$ to $1$), the operator destroys the left mover coming from the origin. This is done with the following non-zero components (where we use again the notation $b=(b_1,b_2)$),
		\begin{eqnarray*}
			j=2p+\frac{1}{2} , p \in \mathbb{Z} \quad ({\rm left \; mover}): &\quad & \mathcal{W} [j]^{False, (2,0)}_{True, (1,b)} =  \mathcal{T}\cdot \mathcal{U}^{\frac{1}{2}}[j]_{False, b}^{False, (0,0)}  \\
			j=2p , p \in \mathbb{Z} \quad ({\rm splitting \; pair}): &\quad & \mathcal{W} [j]^{False, (2,0)}_{True, (1,b)} = \mathcal{T}\cdot \mathcal{U}^{\frac{1}{2}}[j]_{True, b}^{False, (0,1)} \\
			j=2p +1 , p \in \mathbb{Z}  \quad ({\rm fusing \; pair}) : &\quad & \mathcal{W} [j]^{False, (2,0)}_{True, (1,b)} = \mathcal{T}\cdot \mathcal{U}^{\frac{1}{2}}[j]_{False, b}^{False, (1,0)} .
		\end{eqnarray*}
		Between $x_1$ and $x_2$, the activation index is $1$, and $W_{x_1,x_2}$ must shift the configuration. This is done with the non-zero components
		\begin{eqnarray*}
			\mathcal{W} [j]^{\sigma', (1,b')}_{\sigma, (1,b)} = \mathcal{T}\cdot \mathcal{U}^{\frac{1}{2}}[j]_{\sigma, b}^{\sigma', b'} .
		\end{eqnarray*}
		At $x_2$, the activation index goes from $1$ to $0$; the corresponding non-zero components are
		\begin{eqnarray*}
			\mathcal{W} [j]^{True, (1,b')}_{True, (0,0)} =1,
		\end{eqnarray*}
		for all $j$ and all $b'$. Finally, on the right of $x_2$ (activation index $0$), $W_{x_1,x_2}$ acts as the identity, and the corresponding non-zero components are
		\begin{eqnarray*}
			&& \mathcal{W} [j]^{\sigma, (0,0)}_{\sigma, (0,0)} = 1 .
		\end{eqnarray*}

		\subsection{Fourth term in Eq. (\ref{eq:app:nondiag}): $M U^{-t} \left| 01\check{1}0 \right> \left<0 1\check{0} 0 \right| U^t M^\dagger$}	
		
		This term is related to the second one (section \ref{sec:app:case2}) by the reflection $j \rightarrow -j$, which is a symmetry of the model.

		\subsection{Fifth term in Eq. (\ref{eq:app:nondiag}): $M U^{-t} \left| 11\check{1}0 \right> \left<1 1\check{0} 0 \right| U^t M^\dagger$}	
		
		This term is related to the third one (section \ref{sec:app:case3}) by reflection $j \rightarrow -j$.

		\subsection{Sixth term in Eq. (\ref{eq:app:nondiag}): $M U^{-t} \left| 01\check{1}10 \right> \left<0 1\check{0} 1 0 \right| U^t M^\dagger$}	
		
		We write
		$$
		M U^{-t} \left| 01\check{1}10 \right> \left<0 1\check{0} 1 0 \right| U^t M^\dagger = \sum_{x_1,x_2}  (M U^{-t} \left| 1\check{1}1 \right> \left< 1\check{1} 1  \right| U^t M^\dagger) W_{x_1,x_2},
		$$
		where $ (M U^{-t} \left| 1\check{1}1 \right> \left< 1\check{1} 1  \right| U^t M^\dagger)$ is a diagonal operator already studied in Sec.~\ref{sec:app:diag}, and where $W_{x_1,x_2}$ is now the operator that (see Fig. \ref{fig:app:case6})
		\begin{itemize}
			\item evolves the configuration in the interval $[x_1,x_2]$ backwards by one unit time
			\item creates an additional left mover immediately on the left of $x_1$ (if possible, otherwise it annihilates the configuration)
			\item creates an additional right mover immediately on the right of $x_2$  (if possible, otherwise it annihilates the configuration).
		\end{itemize}
		Clearly, each of these three operations can be done with an MPO with finite bond dimension, therefore the combination of the three is also an MPO with finite bond dimension. To elaborate, we 
		\begin{figure}[h]
			\begin{tikzpicture}
			\draw (0,0) node{\includegraphics[width=0.35\textwidth]{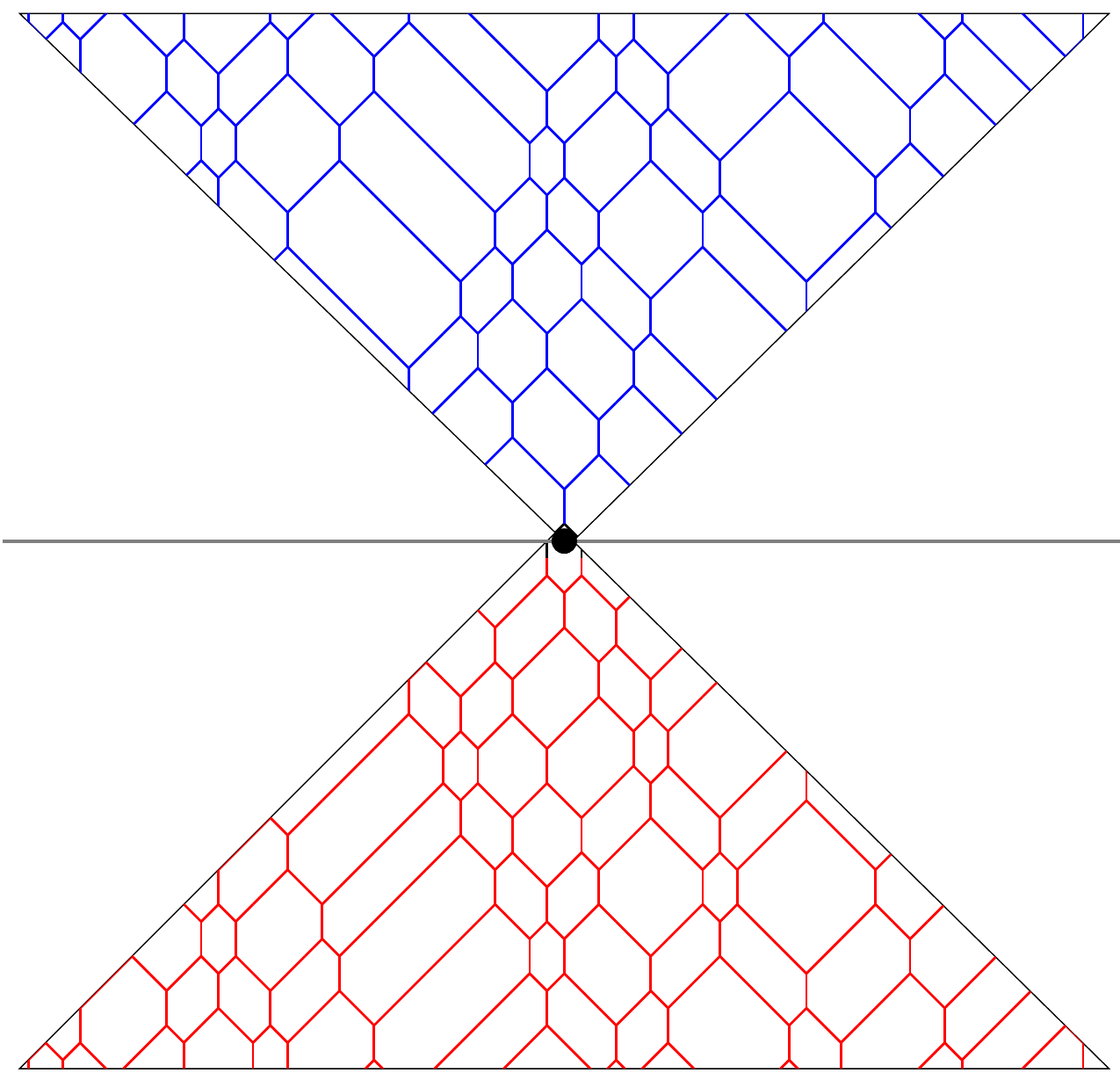}};
			\draw (7,0.4) node{\includegraphics[width=0.4\textwidth]{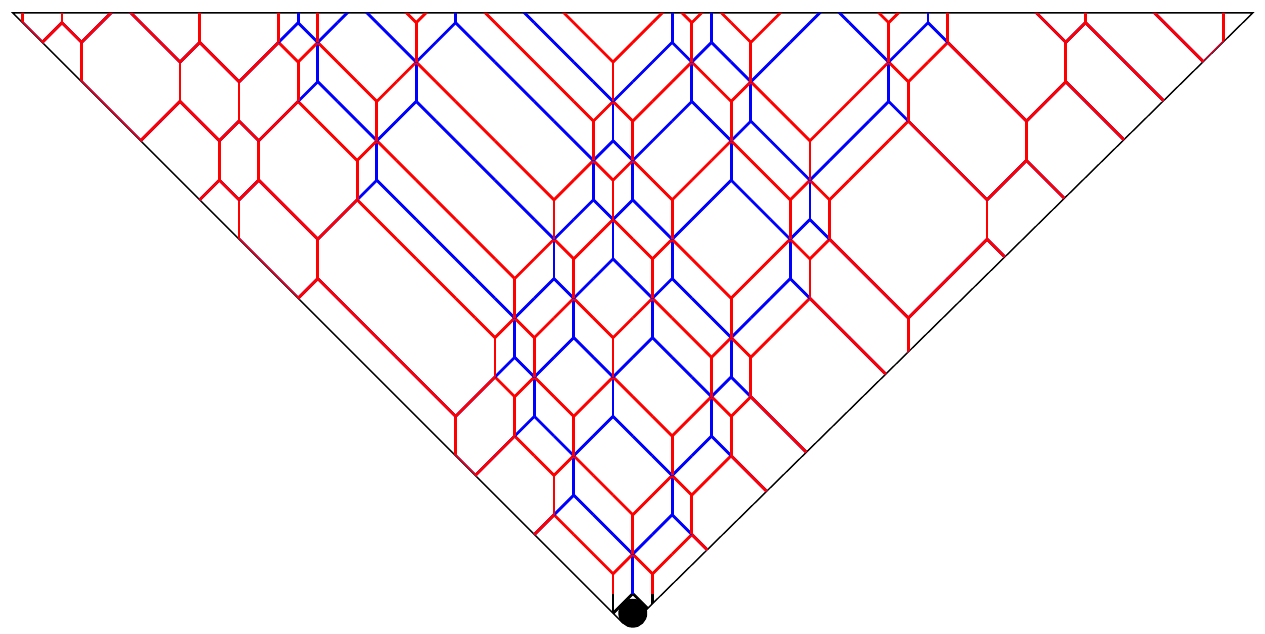}};
			\draw[<-] (5.1,2.2) -- ++(0,0.4) node[above]{$x_1$};
			\draw[<-] (8.68,2.2) -- ++(0,0.4) node[above]{$x_2$};
			%	\draw[<-] (8.12,2.2) -- ++(0,0.4) node[above]{$x_2'$};
			\end{tikzpicture}
			\caption{A typical soliton configuration contributing to $M U^{-t} \left| 01\check{1}10 \right> \left<01 \check{0} 10 \right| U^t M^\dagger$. After folding, the blue configuration is obtained from the red one by applying a time-shift of one half unit time and a translation of one site to the right inside the interval $(x_1,x_2)$ enclosed by the left and right moving soliton that came from the origin.}
			\label{fig:app:case6}
		\end{figure}
		write the MPO for the inverse of the evolution operator with tensors $\mathcal{U}^{-1}[j]^{\sigma', b'}_{\sigma, b}$ whose explicit form is easily adapted from the one of the forward evolution operator, see Sec.~\ref{sec:app:case1}. For positions inside the interval $(x_1,x_2)$, where the activation index is $1$, the non-zero components are
		\begin{eqnarray*}
			\mathcal{W}[j]^{\sigma',(1,b')}_{\sigma,(1,b)} &=& \mathcal{U}^{-1}[j]^{\sigma',b'}_{\sigma,b},
		\end{eqnarray*}
		which takes care of the backward time-shift. Now to add the left mover to the left of $x_1$, we have to distinguish four cases. The first case is when the left mover coming from the origin is at $j = x_1 = 2p + \frac{1}{2}$ (the position where the activation index goes from $2$ to $1$) and there is no right mover at $j-1$.  Then we need to create a new left mover at $j-2 = 2p - \frac{3}{2}$. This is done by chosing the following non-zero components,
		\begin{eqnarray*}
			j=2p-\frac{3}{2} , p \in \mathbb{Z}:  &\quad &  \mathcal{W}[j]^{True,(2,0)}_{False,(2,4)} =  1  , \\
			j=2p-1 , p \in \mathbb{Z}:  &\quad &  \mathcal{W}[j]^{False,(2,4)}_{False,(2,3)} =  1  \\
			j=2p-\frac{1}{2} , p \in \mathbb{Z}:  &\quad &  \mathcal{W}[j]^{False,(2,3)}_{False,(2,2)} =  1 \\
			j=2p , p \in \mathbb{Z}:  &\quad &  \mathcal{W}[j]^{False,(2,2)}_{False,(2,1)} =  1 \\
			j=2p+\frac{1}{2} , p \in \mathbb{Z} \quad ({\rm left \; mover}):  &\quad &  \mathcal{W}[j]^{False,(2,1)}_{True,(1,b)} = \mathcal{U}^{-1}[j]^{False,0}_{True,b} .
		\end{eqnarray*}
		The second case is when the left mover coming from the origin is at $j = x_1 = 2p + \frac{1}{2}$ and there is a right mover at $2p -\frac{1}{2}$. Then the latter needs to be replaced by a pair at $2p-1$. This is done thanks to the additional non-zero components
		\begin{eqnarray*}
			j=2p-1 , p \in \mathbb{Z}:  &\quad &  \mathcal{W}[j]^{True,(2,0)}_{False,(2,5)} =  1  \\
			j=2p-\frac{1}{2} , p \in \mathbb{Z}:  &\quad &  \mathcal{W}[j]^{False,(2,5)}_{True,(2,2)} =  1 .
		\end{eqnarray*}
		The third case is when the left mover coming from the origin is in a fusing pair at $j = x_1=2p+1$, then one has to add a left mover which fuses with the right mover from that pair. Fusing the two gives a new splitting pair at position $x_1-1=2p$. This is implemented by the non-zero components
		\begin{eqnarray*}
			j=2p , p \in \mathbb{Z}:  &\quad &  \mathcal{W}[j]^{True,(2,0)}_{False,(2,7)} =  1 \\
			j=2p +\frac{1}{2}, p \in \mathbb{Z}:  &\quad &  \mathcal{W}[j]^{False,(2,7)}_{False,(2,6)} =  1 \\
			j=2p+1 , p \in \mathbb{Z} \quad ({\rm fusing \; pair}):  &\quad &  \mathcal{W}[j]^{False,(2,6)}_{True,(1,b)} = \mathcal{U}^{-1}[j]^{False,2}_{True,b} .
		\end{eqnarray*}
		The fourth case is when the left mover coming from the origin is in a splitting pair at $j = x_1=2p$, then one must replace it by a left mover at $2p-\frac{3}{2}$ and a right mover at $2p-\frac{1}{2}$. This is achieved by the non-zero components
		\begin{eqnarray*}
			j=2p-\frac{3}{2} , p \in \mathbb{Z}:  &\quad &  \mathcal{W}[j]^{True,(2,0)}_{False,(2,10)} =  1  , \\
			j=2p-1 , p \in \mathbb{Z}:  &\quad &  \mathcal{W}[j]^{False,(2,10)}_{False,(2,9)} =  1  \\
			j=2p-\frac{1}{2} , p \in \mathbb{Z}:  &\quad &  \mathcal{W}[j]^{True,(2,9)}_{False,(2,8)} =  1 \\
			j=2p , p \in \mathbb{Z}  \quad ({\rm splitting \; pair}):  &\quad &  \mathcal{W}[j]^{False,(2,8)}_{True,(1,b)} =  \mathcal{U}^{-1}[j]^{False,2}_{True,b}  .
		\end{eqnarray*}
		This takes care of the addition of the left mover at the left of $x_1$. Apart from this, the operator $W_{x_1,x_2}$ must also act as the identity on the left of $x_1$. This is done by the non-zero components
		$$
		\mathcal{W}[j]^{\sigma,(2,0)}_{\sigma,(2,0)} =  1 ,
		$$
		for all $j$.

		The structure of the components is the same on the right of $x_2$. This leads to the following non-zero components,
		\begin{eqnarray*}
			j=2p+\frac{3}{2} , p \in \mathbb{Z}:  &\quad &  \mathcal{W}[j]^{True,(0,4)}_{False,(0,0)} =  1  , \\
			j=2p+1 , p \in \mathbb{Z}:  &\quad &  \mathcal{W}[j]^{False,(0,3)}_{False,(0,4)} =  1  \\
			j=2p+\frac{1}{2} , p \in \mathbb{Z}:  &\quad &  \mathcal{W}[j]^{False,(0,2)}_{False,(0,3)} =  1 \\
			j=2p , p \in \mathbb{Z}:  &\quad &  \mathcal{W}[j]^{False,(0,1)}_{False,(0,2)} =  1 \\
			j=2p-\frac{1}{2} , p \in \mathbb{Z} \quad ({\rm right \; mover}):  &\quad &  \mathcal{W}[j]^{False,(1,b')}_{True,(0,1)} = \mathcal{U}^{-1}[j]^{False,b'}_{True,0} ,
		\end{eqnarray*}
		\begin{eqnarray*}
			j=2p+1 , p \in \mathbb{Z}:  &\quad &  \mathcal{W}[j]^{True,(0,5)}_{False,(0,0)} =  1  \\
			j=2p+\frac{1}{2} , p \in \mathbb{Z}:  &\quad &  \mathcal{W}[j]^{False,(0,2)}_{True,(0,5)} =  1 ,
		\end{eqnarray*}
		\begin{eqnarray*}
			j=2p , p \in \mathbb{Z}:  &\quad &  \mathcal{W}[j]^{True,(0,7)}_{False,(0,0)} =  1 \\
			j=2p -\frac{1}{2}, p \in \mathbb{Z}:  &\quad &  \mathcal{W}[j]^{False,(0,6)}_{False,(0,7)} =  1 \\
			j=2p-1 , p \in \mathbb{Z} \quad ({\rm fusing \; pair}):  &\quad &  \mathcal{W}[j]^{False,(1,b')}_{True,(0,6)} = \mathcal{U}^{-1}[j]^{False,b'}_{True,1} .
		\end{eqnarray*}
		\begin{eqnarray*}
			j=2p+\frac{3}{2} , p \in \mathbb{Z}:  &\quad &  \mathcal{W}[j]^{True,(0,10)}_{False,(0,0)} =  1  , \\
			j=2p+1 , p \in \mathbb{Z}:  &\quad &  \mathcal{W}[j]^{False,(0,9)}_{False,(0,10)} =  1  \\
			j=2p+\frac{1}{2} , p \in \mathbb{Z}:  &\quad &  \mathcal{W}[j]^{True,(0,8)}_{False,(0,9)} =  1 \\
			j=2p , p \in \mathbb{Z}  \quad ({\rm splitting \; pair}):  &\quad &  \mathcal{W}[j]^{False,(1,b')}_{True,(0,8)} =  \mathcal{U}^{-1}[j]^{False,b'}_{True,1}  ,
		\end{eqnarray*}
		and finally
		$$
		\mathcal{W}[j]^{\sigma,(0,0)}_{\sigma,(0,0)} =  1 
		$$
		for all $j$.

		\subsection{Seventh term in Eq. (\ref{eq:app:nondiag}): $M U^{-t} \left| 01\check{1}11 \right> \left<0 1\check{0} 1 1 \right| U^t M^\dagger$}	
		\label{sec:app:case7}
		
		\begin{figure}[h]
			\begin{tikzpicture}
			\draw (0,0) node{\includegraphics[width=0.35\textwidth]{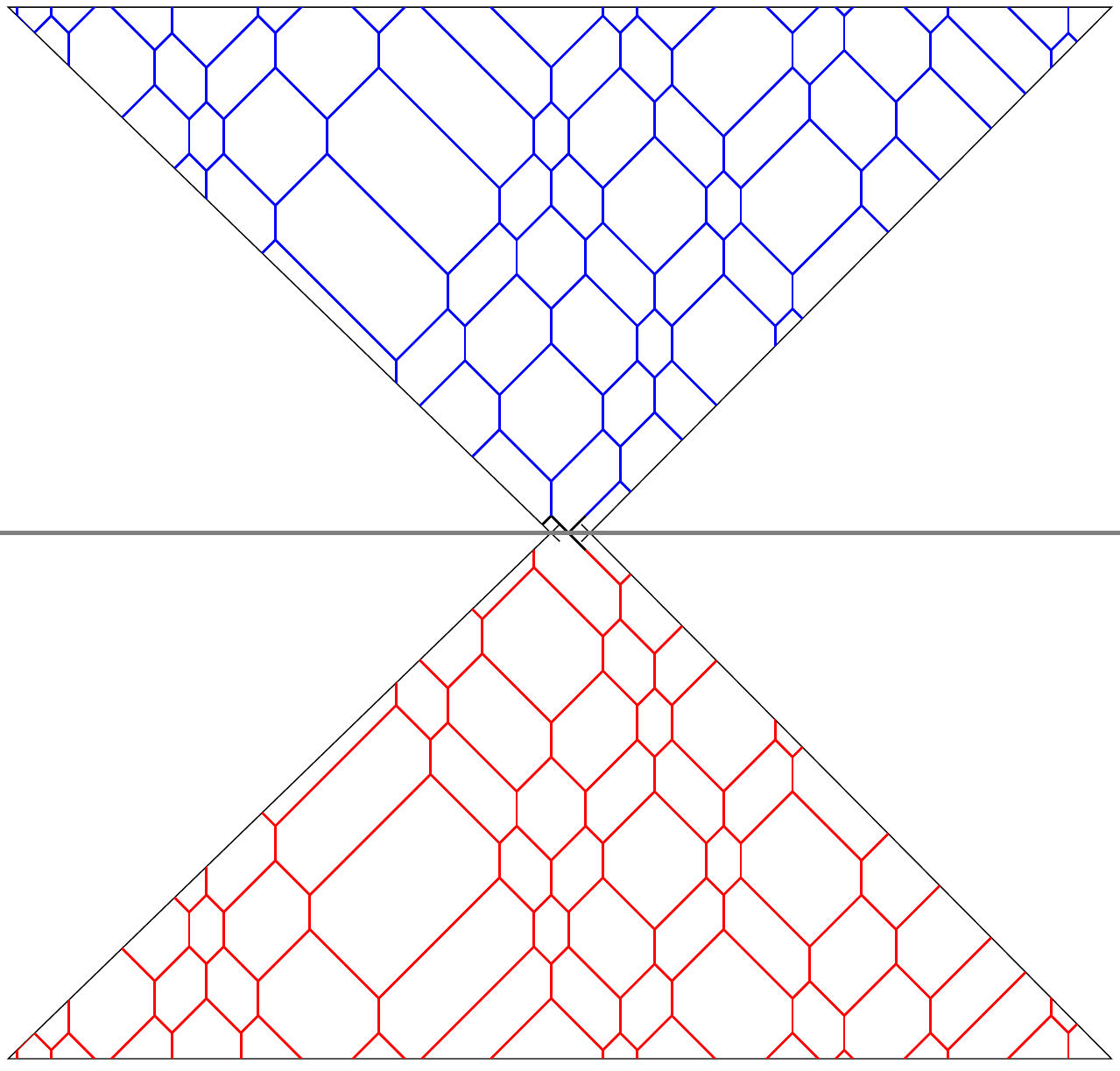}};
			\draw (7,0.4) node{\includegraphics[width=0.4\textwidth]{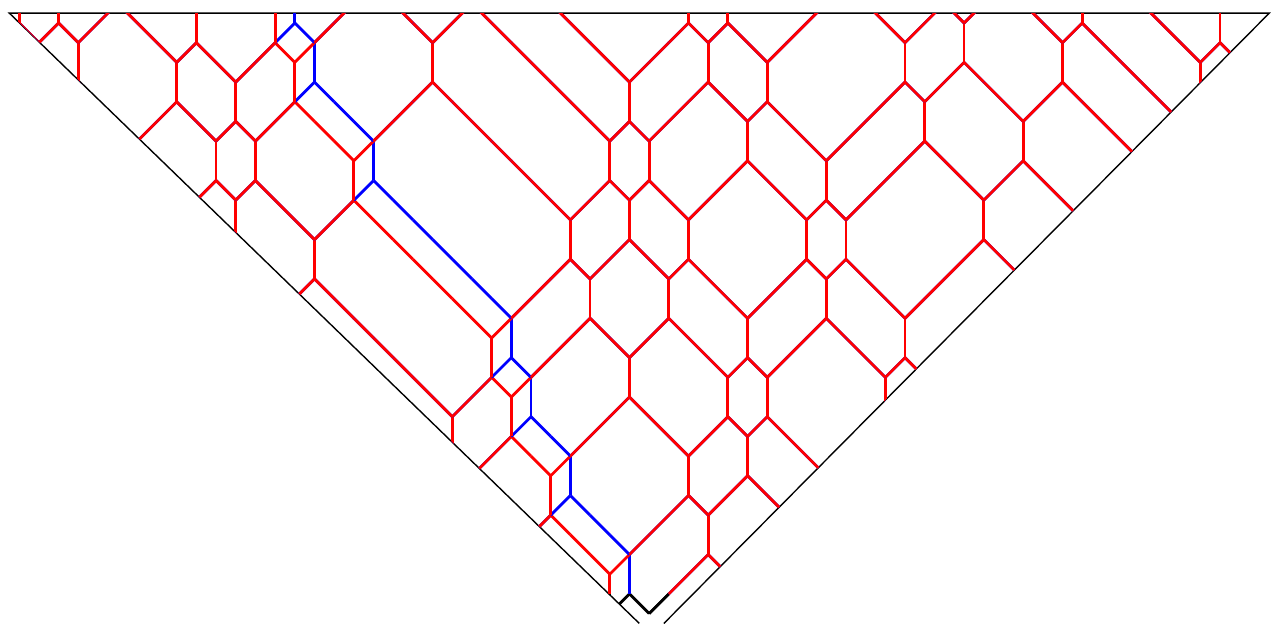}};
			\draw[<-] (5.1,2.2) -- ++(0,0.4) node[above]{$x_1$};
			\draw[<-] (8.61,2.2) -- ++(0,0.4) node[above]{$x_2$};
			\draw[<-] (6.95,-1.35) -- ++(0,-0.3) node[below]{$0$};
			%	\draw[<-] (8.12,2.2) -- ++(0,0.4) node[above]{$x_2'$};
			\end{tikzpicture}
			\caption{A typical soliton configuration contributing to $M U^{-t} \left| 01\check{1}11 \right> \left<01 \check{0} 11 \right| U^t M^\dagger$. After folding, one sees that the red configuration is obtained from the blue one simply by shifting the position of the outgoing left mover (at $x_1$). Notice that there is also a constraint around position $x_2$: all configurations that contribute must have an additional right mover immediately on the right of the one at $x_2$, in order to ensure that there were two right movers at $t=0$: one at $j = -\frac{1}{2}$ and another at $j = \frac{3}{2}$.}
			\label{fig:app:case7}
		\end{figure}
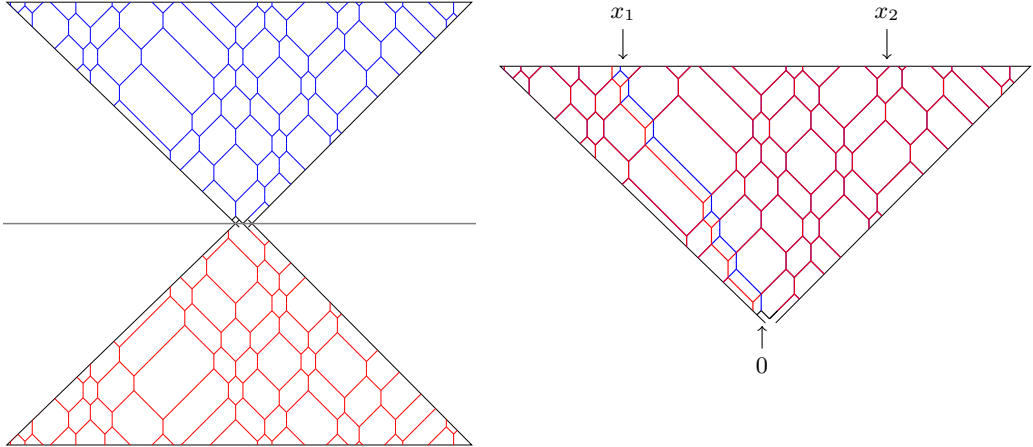
		
		We write this term as
		$$
		M U^{-t} \left| 01\check{1}11 \right> \left<0 1\check{0} 1 1 \right| U^t M^\dagger = \sum_{x_1,x_2}  (M U^{-t} \left| 1\check{1}1 \right> \left< 1\check{1} 1 \right| U^t M^\dagger) W_{x_1,x_2},
		$$
		where $M U^{-t} \left| 1\check{1}1 \right> \left< 1\check{1} 1  \right| U^t M^\dagger$ is the diagonal operator already studied in Sec.~\ref{sec:app:diag}, and where $W_{x_1,x_2}$ is the operator which (see Fig. \ref{fig:app:case7})
		\begin{itemize}
			\item projects onto configurations where there is a right mover immediately to the right of $x_2$
			\item creates a left mover immediately on the left of $x_1$ (if this is possible; if not, then the configuration gets an amplitude zero) and destroys the one at $x_1$.
		\end{itemize}
		Again, each of these operations can be done with an MPO with finite bond dimension, therefore $W_{x_1, x_2}$ has again the same structure as above. We now list the non-zero components.

		On the left of $x_1$, the activation index is $2$ and we have the non-zero components
		\begin{eqnarray*}
			\mathcal{W}[j]^{\sigma,(2,0)}_{\sigma,(2,0)} &=& 1  
		\end{eqnarray*}
		which ensure that $W_{x_1,x_2}$ acts as the identity far on the left. However, $W_{x_1,x_2}$ must also erase the left mover at $x_1$ and create a new left mover immediately on its left. There are four cases to distinguish. If the left mover coming from the origin is in a fusing pair at $j = x_1=2p+1$, then one has to add a left mover which fuses with the right mover from that pair. Fusing the two gives a new splitting pair at position $j-1=x_1-1=2p$. This is implemented by the non-zero components
		\begin{eqnarray*}
			j=2p+1 , p \in \mathbb{Z} \quad ({\rm fusing \; pair}):  &\quad &  \mathcal{W}[j]^{False,(2,1)}_{True,(1,0)} = 1 \\
			j=2p +\frac{1}{2}, p \in \mathbb{Z}:  &\quad &  \mathcal{W}[j]^{False,(2,2)}_{False,(2,1)} =  1 \\
			j=2p , p \in \mathbb{Z}:  &\quad &  \mathcal{W}[j]^{True,(2,0)}_{False,(2,2)} =  1 .
		\end{eqnarray*}
		If the left mover coming from the origin is at $j = x_1 = 2p + \frac{1}{2}$ and there is no right mover at $j-1 = x_1-1$, then we simply have to recreate it at $j-2=x_1-2 = 2p - \frac{3}{2}$. This is done with the non-zero components
		\begin{eqnarray*}
			j=2p+\frac{1}{2} , p \in \mathbb{Z} \quad ({\rm left \; mover}):  &\quad &  \mathcal{W}[j]^{False,(2,3)}_{True,(1,0)} = 1 \\
			j=2p , p \in \mathbb{Z}:  &\quad &  \mathcal{W}[j]^{False,(2,4)}_{False,(2,3)} =  1 \\
			j=2p-\frac{1}{2} , p \in \mathbb{Z}:  &\quad &  \mathcal{W}[j]^{False,(2,5)}_{False,(2,4)} =  1 \\
			j=2p-1 , p \in \mathbb{Z}:  &\quad &  \mathcal{W}[j]^{False,(2,6)}_{False,(2,5)} =  1 \\
			\	j=2p-\frac{3}{2} , p \in \mathbb{Z}:  &\quad &  \mathcal{W}[j]^{True,(2,0)}_{False,(2,6)} =  1 .
		\end{eqnarray*}
		If the left mover coming from the origin is at $j = x_1 = 2p + \frac{1}{2}$ and there is a right mover at $j-1 = x_1-1$, then the latter needs to be replaced by a pair at $x_1-\frac{3}{2}$. This is done with the additional non-zero components
		\begin{eqnarray*}
			%	j=2p-\frac{1}{2} , p \in \mathbb{Z}:  &\quad &  \mathcal{W}[j]^{True,(2,5)}_{False,(2,8)} =  1 \\
			%	j=2p+\frac{1}{2} , p \in \mathbb{Z} \quad ({\rm left \; mover}):  &\quad &  \mathcal{W}[j]^{False,1}_{True,0} = 1 \\
			%	j=2p , p \in \mathbb{Z}:  &\quad &  \mathcal{W}_{.<x_1}[j]^{False,2}_{False,1} =  1 \\
			j=2p-\frac{1}{2} , p \in \mathbb{Z}:  &\quad &  \mathcal{W}[j]^{False,(2,7)}_{True,(2,4)} =  1 \\
			j=2p-1 , p \in \mathbb{Z}:  &\quad &  W[j]^{True,(2,0)}_{False,(2,7)} =  1.
		\end{eqnarray*}
		If the left mover is in a splitting pair at $x_1=2p$, then it must be replaced by a right mover at $x_1-\frac{1}{2}$ and a left mover at $x_1 - \frac{3}{2}$. This is done by the additional non-zero components
		\begin{eqnarray*}
			j=2p , p \in \mathbb{Z} \quad ({\rm splitting \; pair}):  &\quad &  \mathcal{W}[j]^{False,(2,8)}_{True,(1,0)} = 1 \\
			j=2p -\frac{1}{2}, p \in \mathbb{Z}:  &\quad &  \mathcal{W}[j]^{True,(2,5)}_{False,(2,8)} =  1 .%\\
			%	j=2p -1, p \in \mathbb{Z}:  &\quad &  \mathcal{W}[j]^{False,3}_{False,2} =  1 \\
			%	j=2p -\frac{3}{2}, p \in \mathbb{Z}:  &\quad &  \mathcal{W}[j]^{True,0}_{False,3} =  1 .
		\end{eqnarray*}
		Then between $x_1$ and $x_2$ (i.e. where the activation index is $1$) the operator $W_{x_1,x_2}$ acts as the identity. The corresponding non-zero components are
		$$
		\mathcal{W}[j]^{\sigma,(1,0)}_{\sigma,(1,0)}= 1  .
		$$

		Now we need to check that there is a right mover immediately to the right of $x_2$, and there are again a few different cases that need to be distinguished.
		
		If the right mover is in a fusing pair, i.e. if $x_2 = 2 p-1$, then there are two possibilities: there can be either another pair at $x_2 + 2$ or a right mover at $x_2 + \frac{5}{2}$. This is implemented with
		\begin{eqnarray*}
			j=2p-1, p \in \mathbb{Z} \quad ({\rm splitting \; pair}):  &\quad &  \mathcal{W}[j]^{True,(1,0)}_{True,(0,1)} =  \mathcal{W}[j]^{False,(0,4)}_{False,(0,5)}  =  \mathcal{W}[j]^{True,(0,4)}_{True,(0,0)}   = 1 \\
			j=2p-\frac{1}{2} , p \in \mathbb{Z}:  &\quad &  \mathcal{W}[j]^{False,(0,1)}_{False,(0,2)}=  \mathcal{W}[j]^{True,(0,5)}_{True,(0,0)} =   1 \\
			j=2p , p \in \mathbb{Z}:  &\quad &  \mathcal{W}[j]^{False,(0,2)}_{False,(0,3)} =  1 \\
			j=2p +\frac{1}{2}, p \in \mathbb{Z}:  &\quad &  W[j]^{False,(0,3)}_{False,(0,4)} =  1 .
		\end{eqnarray*}
		If the right mover at $x_2$ is not in a pair, i.e. if $x_2 = 2p-\frac{1}{2}$, then there are four acceptable possibilities: either there is a right mover at $x_2+4$ and no soliton between $x_2$ and $x_2+4$, or there is a left mover at $x_2+1 $ and a right mover at $x_2+2 $, or there is a splitting pair at $x_2+\frac{5}{2}$, or there is a fusing pair at $x_2+\frac{7}{2}$. These cases are implemented with the non-zero components
		\begin{eqnarray*}
			j=2p- \frac{1}{2} , p \in \mathbb{Z} \quad ({\rm right \; mover}):  &\quad &  \mathcal{W}[j]^{True,(1,0)}_{True,(0,6)} =  \mathcal{W}[j]^{False,(0,9)}_{False,(0,10)}  =  \mathcal{W}[j]^{True,(0,13)}_{True,(0,0)}   = 1 \\
			j=2p , p \in \mathbb{Z}:  &\quad &  \mathcal{W}[j]^{False,(0,6)}_{False,(0,7)}=  \mathcal{W}[j]^{False,(0,10)}_{False,(0,11)}=   \mathcal{W}[j]^{True,(0,10)}_{True,(0,0)}=  1 \\
			j=2p +\frac{1}{2}, p \in \mathbb{Z}:  &\quad &  \mathcal{W}[j]^{False,(0,7)}_{False,(0,8)}=   \mathcal{W}[j]^{False,(0,11)}_{False,(0,12)} =  \mathcal{W}[j]^{True,(0,7)}_{True,(0,12)}  = 1 \\
			j=2p +1, p \in \mathbb{Z}:  &\quad &  W[j]^{False,(0,8)}_{False,(0,9)} =  W[j]^{False,(0,12)}_{False,(0,13)} =   W[j]^{True,(0,12)}_{True,(0,0)}= 1 .
		\end{eqnarray*}
		If the right mover is in a splitting pair, i.e. if $x_2 = 2 p$, then there are three possibilities: there can be another pair either at $x_2 + 2$ or at $x_2+3$, or there can be a right mover at $x_2 + \frac{7}{2}$. This is implemented with
		\begin{eqnarray*}
			j=2p, p \in \mathbb{Z} \quad ({\rm splitting \; pair}):  &\quad &  \mathcal{W}[j]^{True,(1,0)}_{True,(0,14)} =  \mathcal{W}[j]^{False,(0,17)}_{False,(0,18)}  =  \mathcal{W}[j]^{True,(0,17)}_{True,(0,0)}   = 1 \\
			j=2p+\frac{1}{2} , p \in \mathbb{Z}:  &\quad &  \mathcal{W}[j]^{False,(0,14)}_{False,(0,15)}=  \mathcal{W}[j]^{False,(0,18)}_{False,(0,19)} =   1 \\
			j=2p +1, p \in \mathbb{Z}:  &\quad &  \mathcal{W}[j]^{False,(0,15)}_{False,(0,16)} =   \mathcal{W}[j]^{False,(0,19)}_{False,(0,20)} = \mathcal{W}[j]^{True,(0,19)}_{True,(0,0)} = 1 \\
			j=2p +\frac{3}{2}, p \in \mathbb{Z}:  &\quad &  W[j]^{False,(0,16)}_{False,(0,17)} =  W[j]^{True,(0,20)}_{True,(0,0)}= 1 .
		\end{eqnarray*}
		Finally, further on the right of $x_2$, $W_{x_1,x_2}$ again acts as the identity, and the corresponding non-zero components are
		$$
		\mathcal{W}[j]^{\sigma,(0,0)}_{\sigma,(0,0)}= 1  
		$$
		for all $j$.

		\subsection{Eighth term in Eq. (\ref{eq:app:nondiag}): $M U^{-t} \left| 11\check{1}10 \right> \left<1 1\check{0} 10 \right| U^t M^\dagger$}	
		
		This term is related to the seventh one (section \ref{sec:app:case7}) by reflection $j \rightarrow -j$.
		
		\subsection{Ninth term in Eq. (\ref{eq:app:nondiag}): $M U^{-t} \left| 11\check{1}11 \right> \left<1 1\check{0} 11 \right| U^t M^\dagger$}

		\begin{figure}[h]
			\begin{tikzpicture}
			\draw (0,0) node{\includegraphics[width=0.35\textwidth]{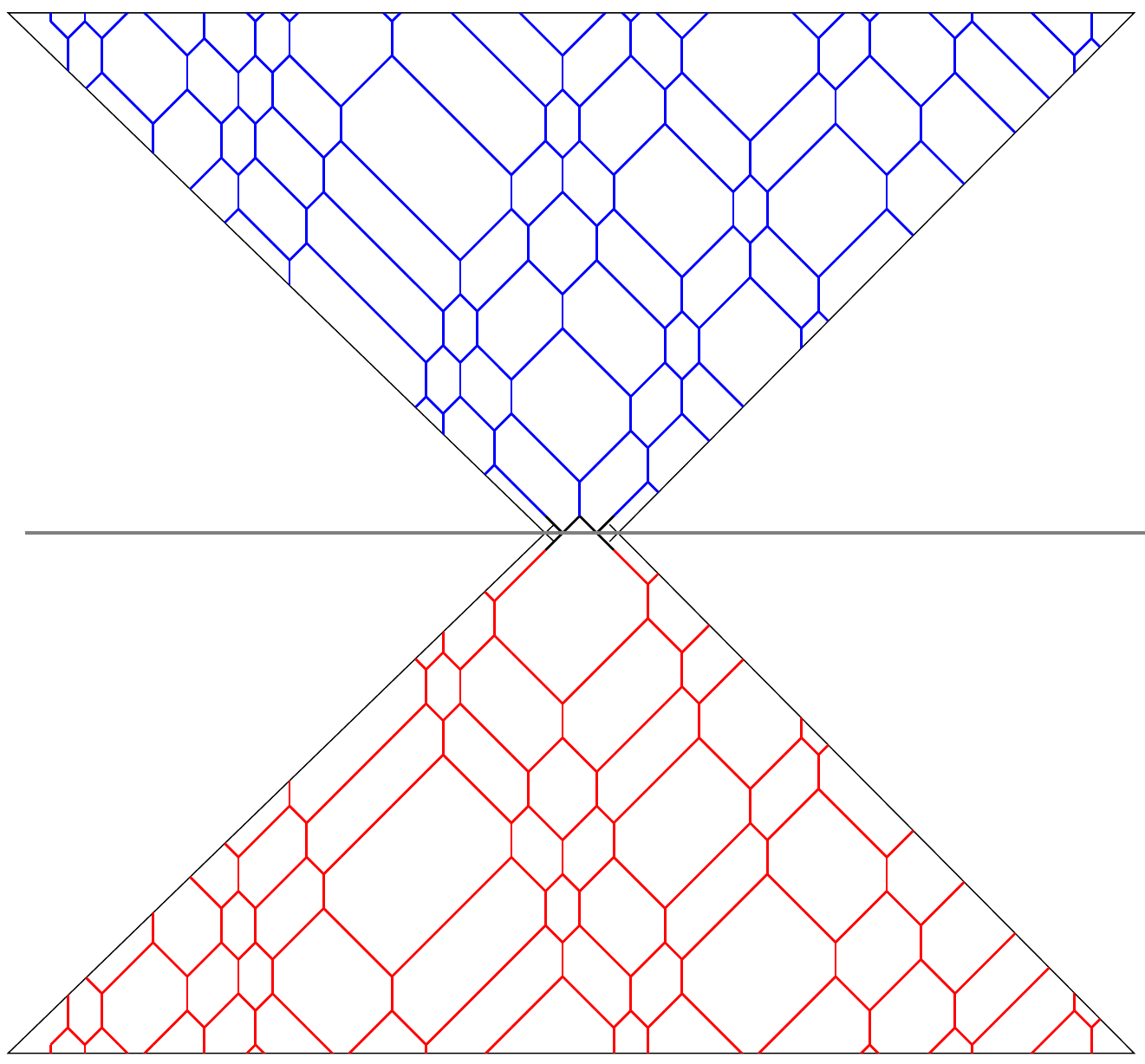}};
			\draw (7,0.4) node{\includegraphics[width=0.4\textwidth]{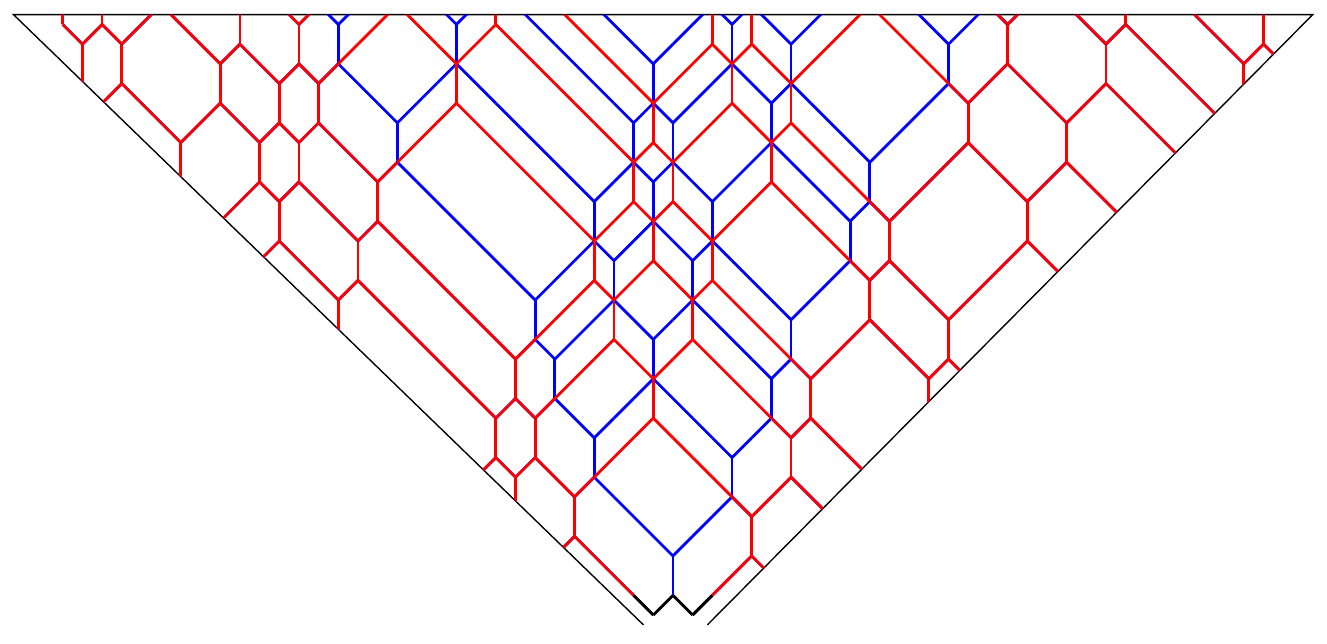}};
			\draw[<-] (5.21,2.2) -- ++(0,0.4) node[above]{$x_1$};
			\draw[<-] (8.68,2.2) -- ++(0,0.4) node[above]{$x_2$};
			%	\draw[<-] (6.95,-1.35) -- ++(0,-0.3) node[below]{$0$};
			%	\draw[<-] (8.12,2.2) -- ++(0,0.4) node[above]{$x_2'$};
			\end{tikzpicture}
			\caption{A typical soliton configuration contributing to $M U^{-t} \left| 11\check{1}11 \right> \left<11 \check{0} 11 \right| U^t M^\dagger$. After folding, one sees that the red configuration is obtained from the blue one simply by shifting the position of the outgoing left mover (at $x_1$). Notice that there is also a constraint around position $x_2$: all configurations that contribute must have an additional right mover immediately on the right of the one at $x_2$, in order to ensure that there were two right movers at $t=0$: one at $j = -\frac{1}{2}$ and another at $j = \frac{3}{2}$.}
			\label{fig:app:case9}
		\end{figure}

		We write this term as
		$$
		M U^{-t} \left| 11\check{1}11 \right> \left<1 1\check{0} 11 \right| U^t M^\dagger = \sum_{x_1,x_2} (\left| 1\check{1}1 \right> \left< 1\check{0} 1 \right|) W_{x_1,x_2} .
		$$
		The diagonal operator $\left| 1\check{1}1 \right> \left< 1\check{0} 1 \right|$ was studied in Sec.~\ref{sec:app:diag}, and the operator $W_{x_1,x_2}$ acts as follows (see Fig.~\ref{fig:app:case9}):
		\begin{itemize}
			\item it checks that there is a left mover immediately on the left of $x_1$, and destroys the one which is at $x_1$
			\item it checks that there is a right mover immediately on the right of $x_2$, and destroys the one which is at $x_2$
			\item it applies a time shift of one time unit inside the interval $(x_1,x_2)$.
		\end{itemize}
		All these operations have already been discussed in previous sections, and it is clear that one can write $W_{x_1,x_2}$ as an MPO of the general form discussed above.

\end{widetext}

\end{document}